\documentclass[conference]{IEEEtran}
\usepackage{pifont}
\usepackage{amsfonts}
\usepackage{subfigure}

\usepackage{amssymb}
\usepackage{amsmath}
\usepackage{epsfig}

\usepackage{cite}

\begin{document}
\onecolumn
\newtheorem{theorem}{Theorem}
\newtheorem{corollary}{Corollary}
\newtheorem{conjecture}{Conjecture}
\newtheorem{definition}{Definition}
\newtheorem{lemma}{Lemma}
\newtheorem{algorithm}{Algorithm}
\newtheorem{remark}{Remark}
\newtheorem{idea}{Idea}
\newtheorem{observation}{Observation}

\newcommand*{\QEDB}{\hfill\ensuremath{\square}}%
\newcommand{\define}{\stackrel{\triangle}{=}}

\pagestyle{empty}

\def\sDoF{\overline{\mbox{\normalfont \scriptsize DoF}}}

\def\QED{\mbox{\rule[0pt]{1.5ex}{1.5ex}}}
\def\proof{\noindent{\it Proof: }}
\date{}

\title{On Heterogeneous Coded Distributed Computing}
\author{\IEEEauthorblockN{Mehrdad Kiamari$^*$, Chenwei Wang$^{\dagger}$ and A. Salman Avestimehr$^*$}\\
\IEEEauthorblockA{$^*$Department of Electrical Engineering, University of Southern California, Los Angeles, CA\\
$^{\dagger}$DOCOMO Innovations, Inc., Palo Alto, CA}}
\maketitle

\thispagestyle{empty}
\begin{abstract}
We consider the recently proposed Coded Distributed Computing (CDC) framework \cite{LMA_all, Avestimehr_CDC, li2016fundamental} that leverages carefully designed redundant computations to enable coding opportunities that substantially reduce the communication load of distributed computing. We generalize this framework to heterogeneous systems where different nodes in the computing cluster can have different storage (or processing) capabilities. We provide the information-theoretically optimal data set placement and coded data shuffling scheme that minimizes the communication load in a cluster with 3 nodes. For clusters with $K>3$ nodes, we provide an algorithm description to generalize our coding ideas to larger networks.

\end{abstract}

\allowdisplaybreaks


\section{Introduction}\label{sec:intro}

The modern paradigm for large-scale distributed computing involves a massively large distributed system comprising individually small and relatively unreliable computing nodes made of commodity low-end hardware.  Specifically, distributed computational frameworks like MapReduce~\cite{dean2004mapreduce}, Spark~\cite{spark}, Dryad~\cite{Dryad}, and CIEL~\cite{Ciel} have gained significant traction, as they enable the execution of production-scale tasks on data sizes of the order of tens of terabytes and more. However, as we ``scale out'' computations across many distributed nodes, massive amounts of raw and (partially) computed data must be moved among nodes, often over many iterations of a running algorithm, to execute the computational tasks. This creates a substantial \emph{communication bottleneck}. For example, by analyzing Hadoop traces from Facebook, it is demonstrated that, on average, 33\% of the overall job execution time is spent on data shuffling~\cite{chowdhury2011managing}. This ratio can be much worse for sorting and other basis tasks underlying many machine learning applications. For example, as shown in~\cite{zhang2013performance}, $50\%\sim70\%$ of the execution time can be spent for data shuffling in applications including TeraSort, WordCount, RankedInvertedIndex, and SelfJoin.

Recently, it has been been shown that ``coding'' can provide novel opportunities to significantly slash the communication load of distributed computing by leveraging carefully designed redundant local computations at the nodes (which can be viewed as creating ``side information''). In particular, a coding framework, named \emph{Coded Distributed Computing (CDC)}, has been proposed in \cite{LMA_all,Avestimehr_CDC,li2016fundamental}
, which assigns the computation of each task at $r$ \emph{carefully chosen} nodes (for some $r \in \mathbb{Z}^+$), in order to enable in-network coding opportunities  that reduce the communication load by  $r$ times. For example, by redundantly computing each task at only \emph{two} carefully chosen nodes, CDC can reduce the communication load by 50\%. The impact of CDC has also been numerically demonstrated  through experiments over Amazon EC2. For example, in~\cite{CTS16} it is shown that in a 16-node cluster, CDC cuts down the execution time of the well-known distributed sorting algorithm \texttt{TeraSort}~\cite{o2008terabyte} by more than $70\%$.

However, CDC have so far been studied and developed for  homogeneous computing clusters. In distributed computing networks different nodes have often different processing, storage, and communication capabilities. For example, Amazon EC2~\cite{AmazonEC2} provides users with a wide selection of instance types with varying combinations of CPU, memory, storage, and bandwidth. Moreover, as discussed in \cite{zaharia2008improving}, the computing environments in virtualized data centers are heterogeneous and algorithms based on homogeneous assumptions can result in significant performance reduction. Our goal in this paper is to take the first steps towards development of CDC for heterogeneous computing clusters. In particular, we aim to understand how we should optimally assign the computation tasks and design optimal coded shuffling techniques in heterogeneous computing clusters.

From homogeneous to heterogeneous, although their CDC developments both rely on creating index coding-type coding opportunities, the problem in heterogeneous systems appears to be much more challenging, due to the fact that we have to deal with more parameters of storage size of nodes for file allocation. In addition, in homogeneous systems, the file allocation to achieve the minimum communication load turned out to be cyclically symmetric with node indices. In contrast, such a manner of file allocation is impossible for heterogeneous systems.

To shed light on CDC for heterogeneous systems, in this paper, we focus on the smallest heterogeneous system with $K\!=\!3$ nodes and characterize the information-theoretically minimum communication load for arbitrary storage size of all nodes. For the achievability, we resolve the main challenge of designing file allocation at each node and then identify how to create coding opportunities on top of carefully designed file allocation. For the converse, we provide a total of four bounds. While two of them are translated directly from \cite{Avestimehr_CDC}, the other two bounds, derived by incorporating genie-aided arguments with the cut-set bounds, are novel. 

For $K\!>\!3$, generalizing the ideas used for developing the information theoretic result for $K=3$ appears to be insufficient due to the fact that when the number of parameters $\{M_k\}_{k=1}^K$ \emph{linearly} grows, the number of possible coding opportunities \emph{exponentially} increases. Specifically, for the achievability, we need to examine if there exists any coding opportunity among every possible subset of $K'$ out of the total $K$ nodes for every $3\leq K'\leq K$. Regarding the converse, investigating whether an achievable scheme is information-theoretically optimal in general is not possible due to lack of efficient tools. Because of these difficulties, we provide a heuristic algorithm to formulate the problem into a linear programming optimization to design the file allocation and the corresponding communication load.

The problem of coded computing in heterogeneous systems has also been studied recently in~\cite{SauravISIT}. However, the focus of that work has been on coded computing approaches that deal with the straggler problem, e.g., \cite{lee2015speeding}, as opposed to the communication load minimization that is the focus of this paper. An interesting future direction is the development of a unified coded computing method for heterogeneous systems that deals with both the bandwidth and straggler problems. Such a unified framework has been proposed for homogeneous systems in \cite{li2016unified}, but remains open for heterogeneous systems.

\section{System Model and Main Results}

\begin{figure}[!t] \centering 
\includegraphics[width=4in]{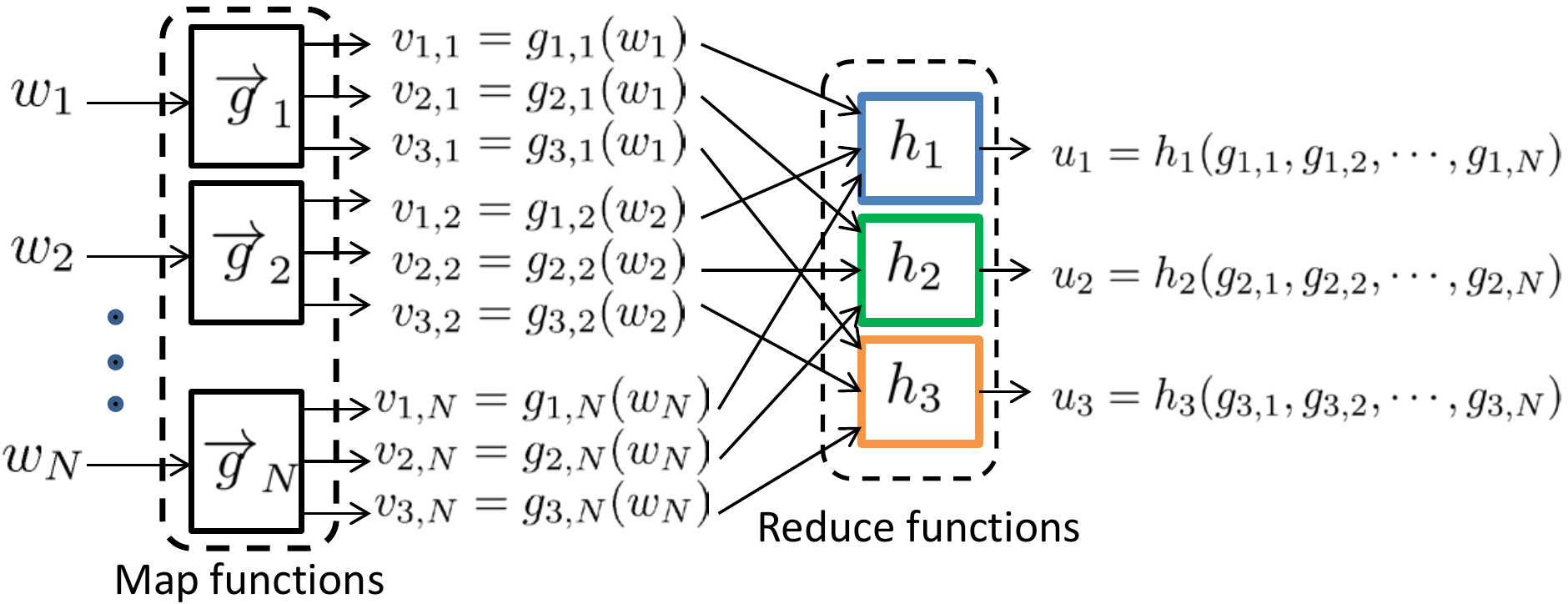}
\caption{A heterogeneous distributed computing system with $Q\!=\!K\!=\!3$}\label{fig:sys_model}
\end{figure}

We consider a \emph{heterogeneous} distributed computing system which consists of $K$ distributed nodes, $N$ input files $\{w_n\}_{n=1}^N$, and each $w_n \in {\mathbb{F}}_{2^F}$ for some $F \in \mathbb{Z}^+$. We assume that each node $k\!\in\!\mathcal{K}\triangleq \{1,\cdots,K\}$ can only store $M_k$ files out of the total $N$ files, i.e., the storage size is $M_k$. In addition, we use $\mathcal{M}_k\!\subseteq\! \{w_1,\cdots,w_N\}$ with the cardinality $M_k=|\mathcal{M}_k|$ to denote the files stored at node $k$. For simplicity, we denote the set of all files by $\mathcal{N}\triangleq \{1,2,\cdots,N\}$, and when there is no ambiguity we simplify the notation to $\mathcal{M}_k\subseteq \mathcal{N}$ to represent files stored at node $k$. In MapReduce-based distributed computing introduced in \cite{dean2004mapreduce}, the goal is to compute $Q$ (for some $Q/K\in \mathbb{Z}^+$) output functions $\phi_1,\dots,\phi_Q$ where each $\phi_q:({\mathbb{F}}_{2^F})^N\rightarrow {\mathbb{F}}_{2^B}$ maps the file $w_n$, $n\in\mathcal{N}$ into an output length-$B$ file $u_q=\phi_q(w_1,\dots,w_N)\in {\mathbb{F}}_{2^B}$ for some $B\in \mathbb{Z}^+$. As depicted in Fig. \ref{fig:sys_model}, the output function $\phi_q$, $\forall q\in\{1,\dots,Q\}$ can be decomposed as follows
\begin{equation}
\phi_q(w_1,\dots,w_N)=h_q(g_{q,1}(w_1),\dots,g_{q,N}(w_N)),
\label{define_MapRed_D}
\end{equation}
where ${\overrightarrow{g}}_n=(g_{1,n},\cdots,g_{Q,n}): {\mathbb{F}}_{2^{F}}\rightarrow ({\mathbb{F}}_{2^T})^Q$, $\forall n\in\mathcal{N}$ for some $T\in \mathbb{Z}^+$, represent the Map functions, and $h_q:({\mathbb{F}_{2^T}})^N\rightarrow {\mathbb{F}_{2^B}}$, $\forall q\in\{1,\dots,Q\}$ represent the Reduce functions. The MapReduce-based distributed computing consists of the following three phases:

\noindent{\bf Map Phase:}  Node $k\in\mathcal{K}$ computes the Map functions on each of the files in $\mathcal{M}_k$ to obtain $Q$ length-$T$ intermediate values computed from each file $w_n$, i.e., $v_{q,n} = g_{q,n}(w_n) \in \mathbb{F}_{2^T}$ for $q=1,\cdots,Q$.

\noindent{\bf Shuffle Phase: } Node $k\in\mathcal{K}$ creates a message $X_k$ which is a function of the intermediate values $\{v_{q,n}|w_n\!\in\!\mathcal{M}_k,q\!\in\!\{1,\cdots,Q\}\}$ locally computed at itself during the Map phase, i.e., $X_k={\psi}_k(\{{\overrightarrow{g}_n}|w_n \in {\mathcal M}_k\})$, and then broadcasts it to all the other nodes.

\noindent{\bf Reduce Phase: } Node $k\in\mathcal{K}$ utilizes the intermediate values computed from its locally stored files during the Map phase 
and the messages $\{X_k\}_{k=1}^K$ collected during the Shuffle phase to recover all its desired intermediate values $\{v_{q,n}|q\!\in\!\mathcal{W}_k,n\!\in\!\mathcal{N}\}$
via the Reduce functions $\{h_q\}_{q\in\mathcal{W}_k}$, where $\mathcal{W}_k$ is the subset of indices of functions interesting to node $k$. Similar to \cite{Avestimehr_CDC}, we assume that $\cup_k \mathcal{W}_k=\{1,\cdots,Q\}$, $\mathcal{W}_i\cap\mathcal{W}_j=\emptyset$ for $\forall i,\forall j,i\neq j$, and $|\mathcal{W}_k|=Q/K$.

The communication load we use in this paper, represented by ${\mathcal L}$, is defined as the total number of bits broadcasted by the $K$ nodes, during the Shuffle phase, normalized by $T$. Intuitively, it means the total number of equations associated with the intermediate values (or files) contributed by the messages $\{X_k\}_{k=1}^K$. To read the notations easily, we illustrate one example in Fig. \ref{fig:sys_model} with $K=3$ and by letting $Q=K$, which means that each node $k$ desires only one intermediate value for each file $v_{k,n},n\in\mathcal{N}$, represented by the blue (node 1), green (node 2) and brown (node 3) colors, respectively. For the file $w_n,n\in\mathcal{N}$, the $k^{th}$ intermediate value desired by node $k$ is denoted as $v_{k,n}$ or $V_{k,n}$ which is a random variable when we study the converse. Since each file has to be stored at one node at least, we also assume that $\cup_k \mathcal{M}_k= \mathcal{N}$ and $M\triangleq \sum_{k}M_k\!\geq\! N$.

In this paper, we are primarily interested in the central question: \emph{given the parameters $(\{M_k\}_{k=1}^K,N)$, what is the minimum communication load ${\mathcal L}^*$}? To address this question, we need to find out the file allocation at each node and the coding scheme for data shuffling to achieve ${\mathcal L}^*$. In particular, we will consider first the setting with $K=3$ and then the setting with $K\!>\!3$. The result is shown as follows.

\subsection{Main Result}

For the problem we defined above, for $K=3$, without loss of generality, we assume that $M_1\leq M_2\leq M_3$. We summarize our new result for $K=3$ in the following theorem.

{\bf Theorem 1}: For the CDC problem defined above, given a general setting with $P\triangleq(M_1,M_2,M_3,N)$ and $M=M_1+M_2+M_3$, the minimum communication load ${\mathcal L}^*$ is given by:
\begin{equation}
{\mathcal L}^*=\left\{\begin{array}{ll} \frac{7}{2}N-\frac{3}{2}M& \textrm{if}~P\in \mathcal{R}_1\cup\mathcal{R}_2\cup\mathcal{R}_3\\
3N-(M_1+M)&\textrm{if}~P\in \mathcal{R}_4\cup\mathcal{R}_5\\
\frac{3}{2}N-\frac{1}{2}M&\textrm{if}~P \in \mathcal{R}_6\\
N-M_1&\textrm{if}~P \in \mathcal{R}_7 \end{array}\right.
\end{equation}
where
\begin{equation}\notag
\begin{aligned}
\mathcal{R}_1&\!=\! \{P~|~M_1\!+\!M_2\leq N,~M_3\leq N\!+\!M_1\!-\!M_2\},\\
\mathcal{R}_2&\!=\! \{P~|~M\leq 2N,~M_1\!+\!M_2> N,M_3\leq N\!+\!M_1\!-\!M_2\},\\
\mathcal{R}_3&\!=\! \{P~|~M\leq 2N,~ M_1\!+\!M_2> N,~N\!+\!M_1\!-\!M_2\geq M_3> 3N\!-\!M_1\!-\!3M_2\},\\
\mathcal{R}_4&\!=\! \{P~|~M_1\!+\!M_2\leq N,~M_3> N\!+\!M_1\!-\!M_2\},\\
\mathcal{R}_5&\!=\! \{P~|~M\leq 2N,~ M_1\!+\!M_2> N,M_3> N\!+\!M_1\!-\!M_2\},\\
\mathcal{R}_6&\!=\! \{P~|~M> 2N,~M_3\leq N\!+\!M_1\!-\!M_2\},\\
\mathcal{R}_7&\!=\! \{P~|~M> 2N,~M_3> N\!+\!M_1\!-\!M_2\}.
\end{aligned}
\end{equation}

{\it Proof:} The proof will be presented in Section \ref{sec:achievability} for the achievability and Section \ref{sec:converse} for the converse, respectively. \hfill\QED

\begin{remark}
Compared to the uncoded scheme where the 3 nodes need a total of $3N-M$ intermediate values in the Shuffle phase, Theorem 1 implies that we can save the communication load by up to $3N-M-{\mathcal L}^*$ by carefully designing the file allocation and the coding scheme.
\end{remark}

\begin{remark}
When $M_1=M_2=M_3$, Theorem 1 reduces to the result specified for the homogeneous system in \cite{Avestimehr_CDC} after normalizing ${\mathcal L}$ by $N$. Meanwhile, it can be seen that the inequality $M_3> N+M_1-M_2$ in $\mathcal{R}_4$, $\mathcal{R}_5$ and $\mathcal{R}_7$ identifies  cases which do not exist in the homogeneous system. In addition, comparison with the homogeneous system in \cite{Avestimehr_CDC} implies that $\mathcal{L}^*$ depends on {\em not only} the computation load (defined as $M/N$ in \cite{Avestimehr_CDC}), {\em but also} the storage size $M_k$ (e.g., $M_1$ in $\mathcal{R}_4$, $\mathcal{R}_5$, $\mathcal{R}_7$).
\end{remark}

The key idea for developing the result in Theorem 1 is to carefully design file allocation over nodes so that we can create coding opportunities for reducing the communication load as many as possible. Meanwhile, since the new result depends on the storage size of each node, it is natural to expect that file allocation to achieve ${\mathcal L}^*$ is non-cyclically symmetric with node indices, as opposed to the manner of cyclically symmetric file allocation to achieve ${\mathcal L}^*$ in the homogeneous system.

For $K\!>\!3$, due to the difficulties of characterizing the information-theoretically optimal result, we provide an achievable scheme in Section \ref{sec:generalK} via developing an algorithm that formulates the achievability into a linear programming problem, followed by several discussions. The main idea behind the achievability is to incorporate scaling up the coding schemes proposed for heterogeneous systems with $K\!=\!3$ with the coding schemes for homogeneous systems developed in \cite{Avestimehr_CDC}.

\section{The Achievability of Theorem 1}\label{sec:achievability}

Before proceeding into the proof in detail, we first provide an overview of CDC for heterogeneous systems and build the key intuitions behind the new result, which can be naturally applied when we design the achievable scheme.

Let us consider an example with $(M_1,M_2,M_3,N)=(6,7,7,12)$ where node $k$ is only interested in collecting its desired $N$ intermediate values $\{v_{k,n}\}_{n\in\mathcal{N}}$ where $k=1,2,3$. Clearly, without creating $\mathcal{L}=16$, since besides the 6, 7, 7 intermediate values computed directly from their own files, respectively, these nodes need another 6, 5, 5 intermediate values, respectively, to complete their reductions.

\begin{figure}[!t] \centering 
\includegraphics[width=3.5in]{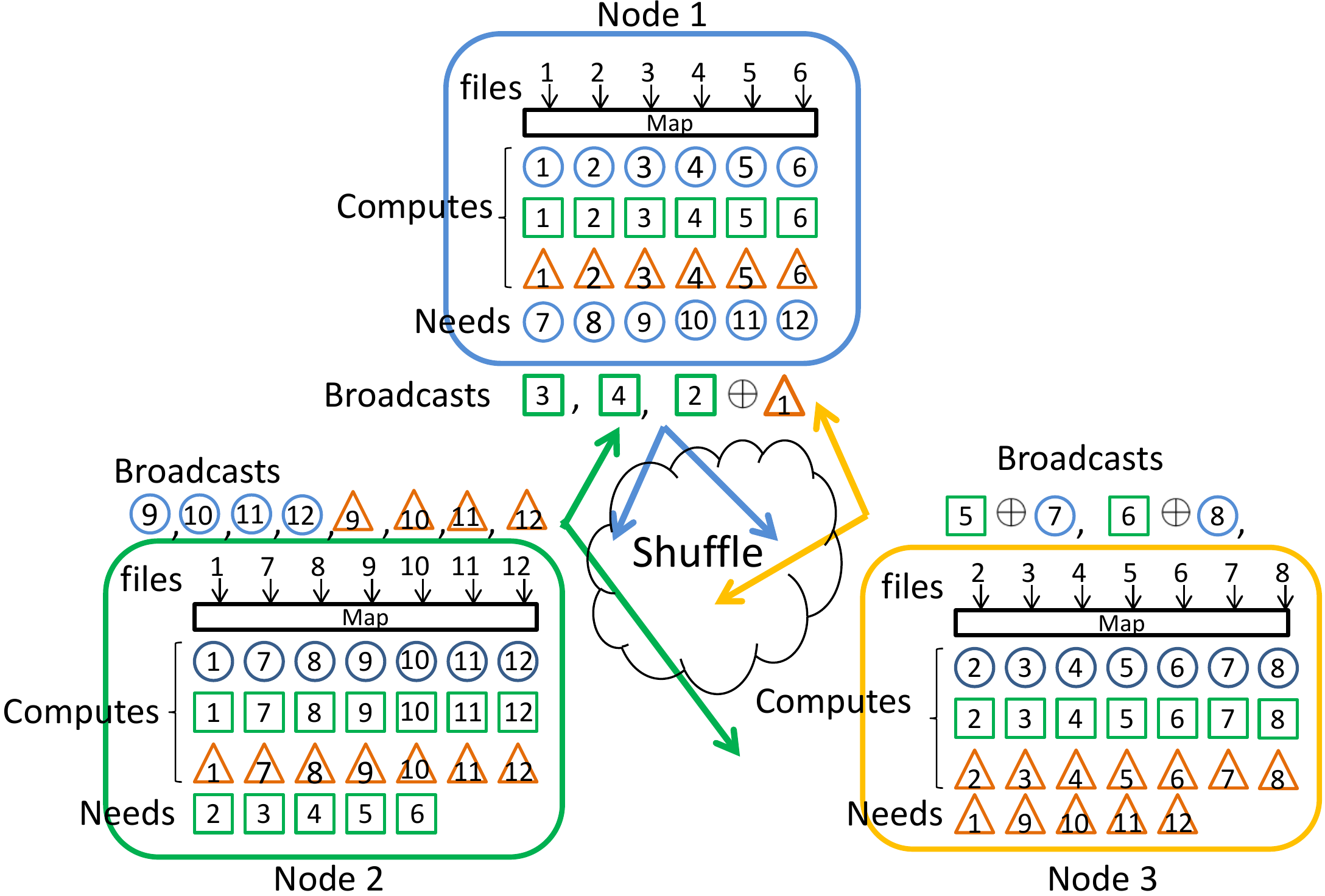}
\caption{An example of non-optimal CDC for heterogeneous network with $(M_1,M_2,M_3,N)=(6,7,7,12)$ and $K=Q=3$ achieves ${\mathcal L}=13$.}\label{coding_ex_non_opt}
\end{figure}

Next, consider the files allocated sequentially as shown in Fig. \ref{coding_ex_non_opt} where node 1 stores files $1-6$, node 2 stores files $7-12$, 1, node 3 stores files $2-8$, and they aim to reduce circle (blue), square (green), and triangle (brown) output functions, respectively. In this case, we can reduce the communication load from $\mathcal{L}=16$ to $\mathcal{L}=13$ by designing an appropriate coding scheme. Specifically, instead of broadcasting the intermediate values $v_{2,2}$ (square 2) and $v_{3,1}$ (triangle 1) individually, node 1 directly broadcasts $v_{2,2}\oplus v_{3,1}$ where $\oplus$ denotes the XOR operator. Since $v_{2,2}$ and $v_{3,1}$ are available at node 3 and node 2, respectively, they can obtain the desired interference-free $v_{3,1}=X_1\oplus v_{2,2}$ and $v_{2,2}=X_1\oplus v_{3,1}$, respectively. Similarly, encoding over 4 intermediate values $v_{2,5}$, $v_{2,6}$, $v_{3,7}$, $v_{3,8}$ at node 3 to $v_{2,5}\oplus v_{3,7}$ and $v_{2,6}\oplus v_{3,8}$ enables saving another 2 transmissions. Thus, we save a total of 3 transmissions.

\begin{figure}[!t] \centering 
\includegraphics[width=3.5in]{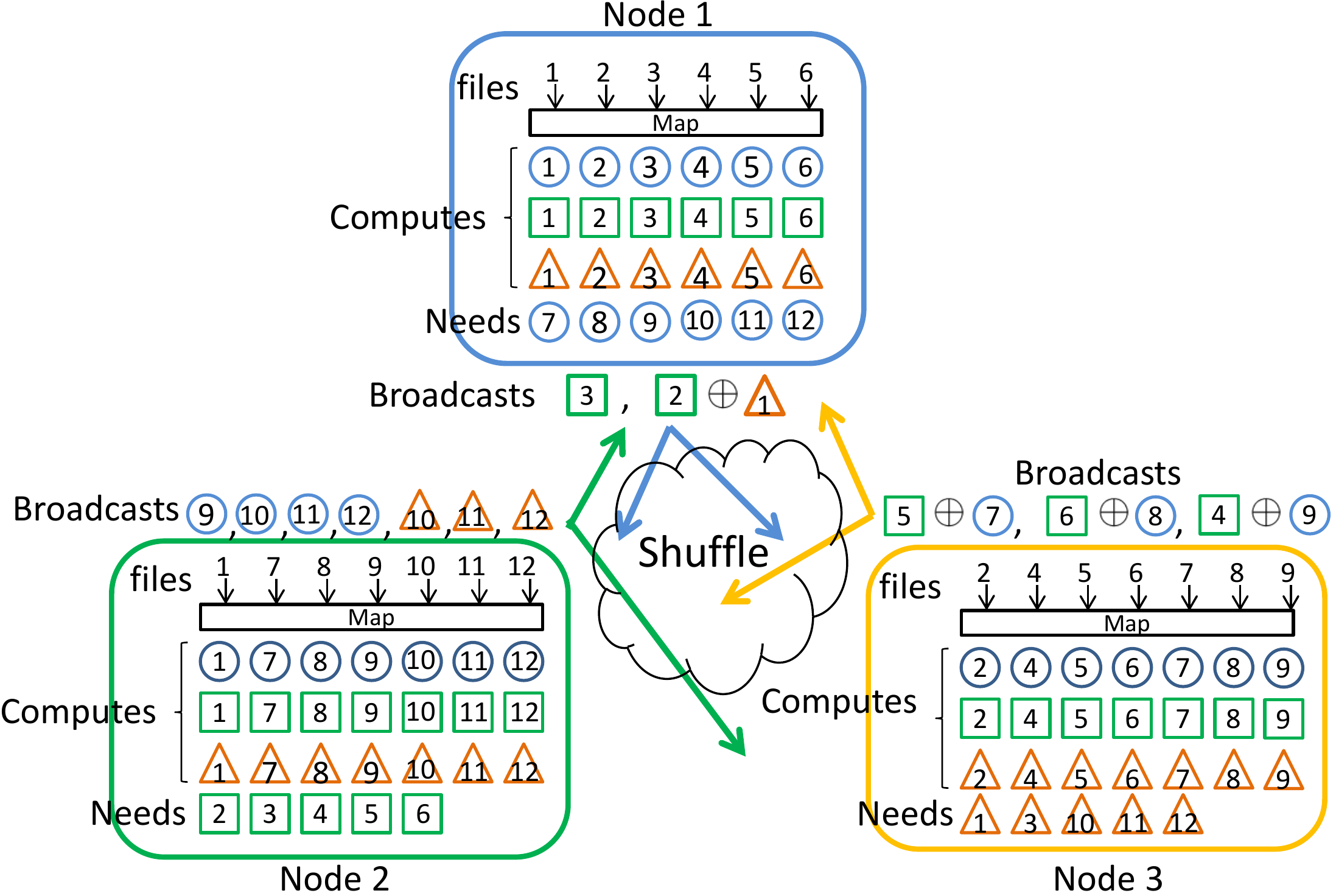}
\caption{An example of optimal CDC for heterogeneous network with $(M_1,M_2,M_3,N)=(6,7,7,12)$ and $K=Q=3$ achieves ${\mathcal L}^*=12$ which is $25\%$ lower compared to uncoded distributed computing.}\label{coding_ex}
\end{figure}

However, for $(M_1,M_2,M_3,N)=(6,7,7,12)$, Theorem 1 implies $\mathcal{L}^*=12$, which means that $\mathcal{L}=13$ shown above is \emph{not} minimum. As depicted in Fig. \ref{coding_ex}, if we carefully design the file allocation at node 3 to be $\mathcal{M}_3=\{2,4,5,6,7,8,9\}$ and keep the file allocation at the other two nodes the same, then we can create another coding opportunity of $v_{2,4}\oplus v_{3,9}$ at node 3. By doing so, we save one more transmission, so as to achieve $\mathcal{L}^*=12$. Meanwhile, as explained later, 12 is also information-theoretically optimal for $(M_1,\!M_2,\!M_3,\!N)\!=\!(6,7,7,12)$ with every possible file allocation.

Based on the example above, it turns out that we need to deal with two coupled challenges including (a) appropriate file allocation over the nodes, and (b) optimal coding scheme design. Let us first consider the challenge (b): given an arbitrary but fixed file allocation, we will show how to create coding opportunities as many as possible. 
In particular, given the file allocation $\mathcal{M}_1$, $\mathcal{M}_2$, $\mathcal{M}_3$, we can always characterize their relationship by identifying the following 7 subsets:
\begin{eqnarray*}
\mathcal{S}_{123}\triangleq\mathcal{M}_1\cap \mathcal{M}_2 \cap\mathcal{M}_3,~~~~\notag\mathcal{S}_{12}&\!\!\!\!\triangleq\!\!\!\!&(\mathcal{M}_1\cap \mathcal{M}_2) \setminus \mathcal{S}_{123},~~~~\mathcal{S}_1\triangleq\mathcal{M}_1 \setminus (\mathcal{M}_2\cup \mathcal{M}_3),\notag\\
\mathcal{S}_{13}&\!\!\!\!\triangleq\!\!\!\!&(\mathcal{M}_1\cap \mathcal{M}_3) \setminus \mathcal{S}_{123},~~~~\mathcal{S}_2\triangleq\mathcal{M}_2 \setminus (\mathcal{M}_1\cup \mathcal{M}_3),\notag\\
\mathcal{S}_{23}&\!\!\!\!\triangleq\!\!\!\!&(\mathcal{M}_2\cap \mathcal{M}_2) \setminus \mathcal{S}_{123},~~~~\mathcal{S}_3\triangleq\mathcal{M}_3 \setminus (\mathcal{M}_1\cup \mathcal{M}_2).
\end{eqnarray*}
For simplicity, we denote the set cardinality by $S_{(\cdot)}\triangleq |\mathcal{S}_{(\cdot)}|$.


Since the subset $\mathcal{S}_{123}$ of files is available at every node, we do not need to consider the communication of the intermediate values $v_{k,\mathcal{S}_{123}}\triangleq \{v_{k,n}\}_{n\in\mathcal{S}_{123}}$ among the 3 nodes. In addition, each of the 3 subsets $\mathcal{S}_1$, $\mathcal{S}_2$ and $\mathcal{S}_3$ of files is available at one node only. Thus, for each node $k$, if the other two nodes want to collect their intermediate values $\{v_{j,\mathcal{S}_k}\}_{j\neq k}$, node $k$'s message $X_k$ has to carry those intermediate values, which need a total of $2(S_1+S_2+S_3)$ transmissions. Hence, the possibility of CDC originates from the remaining 3 subsets $\mathcal{S}_{12}$, $\mathcal{S}_{13}$, $\mathcal{S}_{23}$ where each file is stored at 2 nodes only. With this key observation, we have the following lemma.

{\bf Lemma 1:} Given file allocation $\mathcal{M}\triangleq (\mathcal{M}_1,\mathcal{M}_2,\mathcal{M}_3)$, the communication load ${\mathcal L}_{\mathcal{M}}$ is achievable, where
\begin{eqnarray}
\mathcal{L}_{\mathcal{M}}=2(S_1+S_2+S_3)+g(S_{12},S_{13},S_{23}),\label{eqn:T}
\end{eqnarray}
and
\begin{eqnarray*}
g(x_1,x_2,x_3)=\frac{1}{2}\Big(\Big|\max_k x_k\!+\!\sum_k \frac{x_k}{2}\Big|\!+\!\Big|\max_k x_k\!-\!\sum_k \frac{x_k}{2}\Big|\Big).
\end{eqnarray*}

{\it Proof:} Clearly, we only need to show the achievability of the $g(S_{12},S_{13},S_{23})$ term. In fact, observations of the function $g(S_{12},S_{13},S_{23})$ reveal that if we satisfy the triangle inequality $S_{12}+S_{13}+S_{23}-\max S_{ij}\geq \max S_{ij}$, then we can create a total of $(S_{12}\!+\!S_{13}\!+\!S_{23})/2$ equations. Otherwise, besides the $S_{12}\!+\!S_{13}\!+\!S_{23}\!-\!\max S_{ij}$ equations that we can create at most, we need to send additional $\max S_{ij}\!-(\!S_{12}\!+\!S_{13}\!+\!S_{23}\!-\max S_{ij}\!)$ intermediate values, so the total number of equations is $\max S_{ij}$.
With this intuition, we provide the coding scheme for the two cases, respectively. Without loss of generality, we assume that $S_{12}\leq S_{13}\leq S_{23}$.

\begin{figure}[!b] \centering
\includegraphics[width=3.2in]{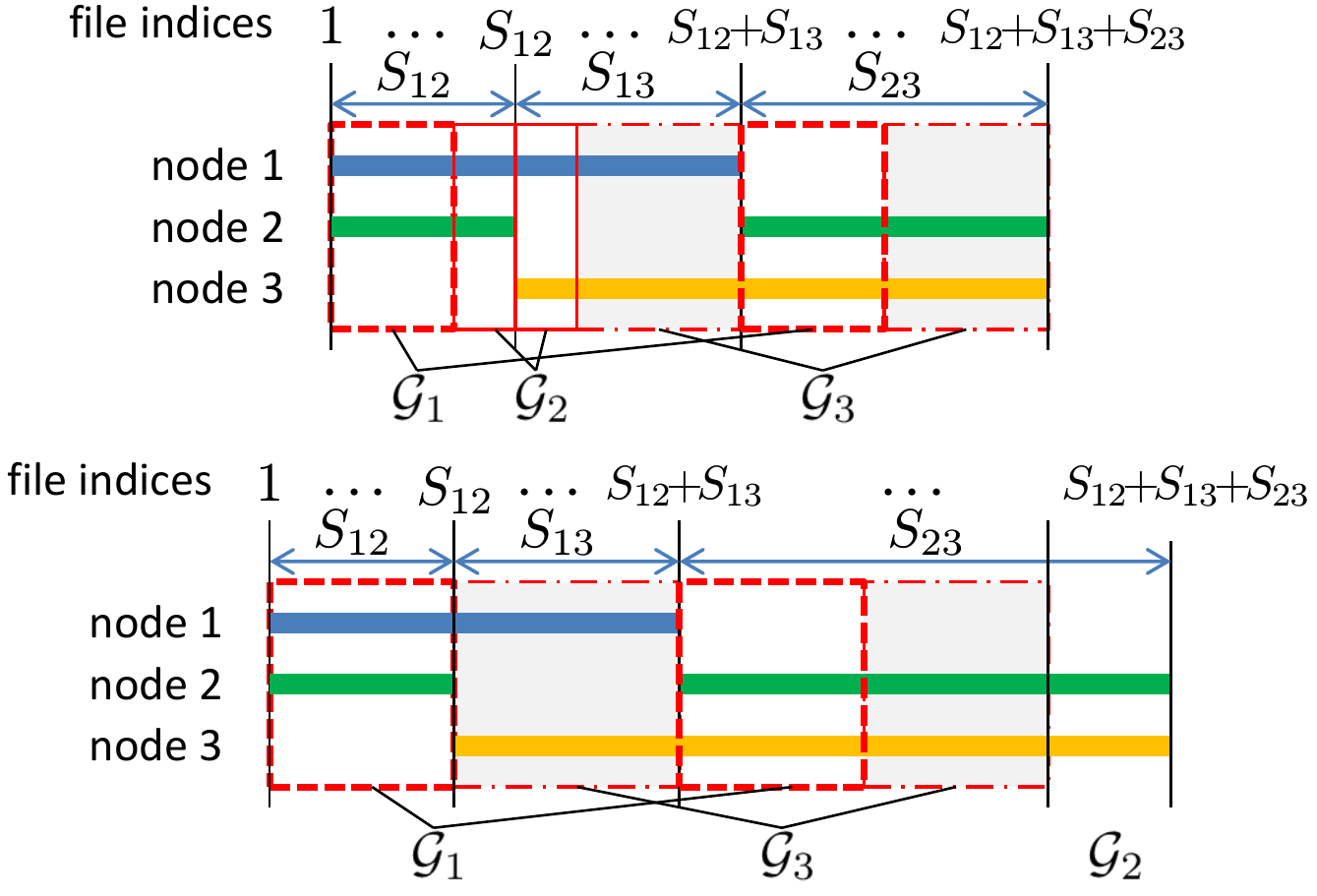}
\caption{Illustration of coding scheme design for (upper) $S_{12}+ S_{13}\geq S_{23}$ and (lower) $S_{12}+ S_{13}< S_{23}$ by considering the files in $\mathcal{S}_{12}$, $\mathcal{S}_{13}$, $\mathcal{S}_{23}$ only}\label{fig:case0}
\end{figure}

{\bf Case 1: $S_{12}+ S_{13}\geq S_{23}\Rightarrow g(S_{12},S_{13},S_{23})=(S_{12}+S_{13}+S_{23})/2$}: As shown in Fig. \ref{fig:case0} (upper), we group the files into 3 non-overlapping groups $\mathcal{G}_l,l=1,2,3$ in 3 rectangles (dashed, solid and mixture), so that each $\mathcal{G}_l$ has the same size of overlapping with two out of the 3 sets $S_{12}$, $S_{13}$, $S_{23}$. 
Denoting the number of files in $\mathcal{G}_l$ as $L_l$, we can resolve $L_l,l=1,2,3$ from the following linear equations:
\begin{eqnarray}
\left\{\begin{array}{l}L_1/2+L_2/2=S_{12}\\L_2/2+L_3/2=S_{13}\\L_1/2+L_3/2=S_{23}\end{array}\right.\Longrightarrow \left\{\begin{array}{l}L_1=S_{12}-S_{13}+S_{23}\\L_2=S_{12}+S_{13}-S_{23}\\L_3=-S_{12}+S_{13}+S_{23},\end{array}\right.
\end{eqnarray}
and we obtain the corresponding files in each group as:
\begin{eqnarray}
\mathcal{G}_1&\!\!\!\!=\!\!\!\!&\mathcal{S}_{12(1:L_1/2)}\cup\mathcal{S}_{23(1:L_1/2)}\\
\mathcal{G}_2&\!\!\!\!=\!\!\!\!&\mathcal{S}_{12(L_1/2+1:S_{12})}\cup\mathcal{S}_{13(1:L_2/2)}\\
\mathcal{G}_3&\!\!\!\!=\!\!\!\!&\mathcal{S}_{13(L_2/2+1:S_{13})}\cup\mathcal{S}_{23(L_1/2+1:S_{23})}.
\end{eqnarray}
Thus, the messages broadcasted from each node are given by:
\begin{eqnarray}
X_1&\!\!\!\!=\!\!\!\!&v_{3,\mathcal{G}_2\cap\mathcal{S}_{12}}\oplus v_{2,\mathcal{G}_2\cap\mathcal{S}_{23}}\\
X_2&\!\!\!\!=\!\!\!\!&v_{3,\mathcal{G}_1\cap\mathcal{S}_{12}}\oplus v_{1,\mathcal{G}_1\cap\mathcal{S}_{23}}\\
X_3&\!\!\!\!=\!\!\!\!&v_{2,\mathcal{G}_3\cap\mathcal{S}_{13}}\oplus v_{1,\mathcal{G}_3\cap\mathcal{S}_{23}}.
\end{eqnarray}
With the design above, it is easy to examine that each node will obtain all desired intermediate values in the shuffling phase.

{\bf Case 2: $S_{12}+ S_{13}< S_{23}\Rightarrow g(S_{12},S_{13},S_{23})=S_{23}$}: With the similar approach, as shown in Fig. \ref{fig:case0} (lower), we split $\mathcal{S}_{23}$ into 3 non-overlapping groups, each with cardinality $2S_{12}$, $2S_{13}$ and $S_{23}-(S_{12}+S_{13})$, respectively. While the former two groups have the same types of file allocation as in $\mathcal{G}_1$ and $\mathcal{G}_3$ in Case 1, respectively, we only need to let node 2 or node 3 directly send the corresponding intermediate values in group 3 (outside of the rectangles in Fig. \ref{fig:case0} (lower)) to node 1. \hfill\QED

So far, given file allocation $\mathcal{M}$, $\mathcal{L}(M)$ can be uniquely determined by Lemma 1. What remains to be shown is the challenge: what file allocation achieves ${\mathcal L}^*$ in Theorem 1. In the rest of this section, we provide the file allocation $\mathcal{M}$ to achieve $\mathcal{L}^*$ for  regimes $\mathcal{R}_1-\mathcal{R}_7$.

\subsection{$M_1+M_2\leq N$}


\subsubsection{$M_3\leq N+M_1-M_2$ (Regime $\mathcal{R}_1$)}

\begin{figure}[!t] \centering 
\includegraphics[trim = 0in 1.7in .5in 2in, clip,width=0.4\textwidth]{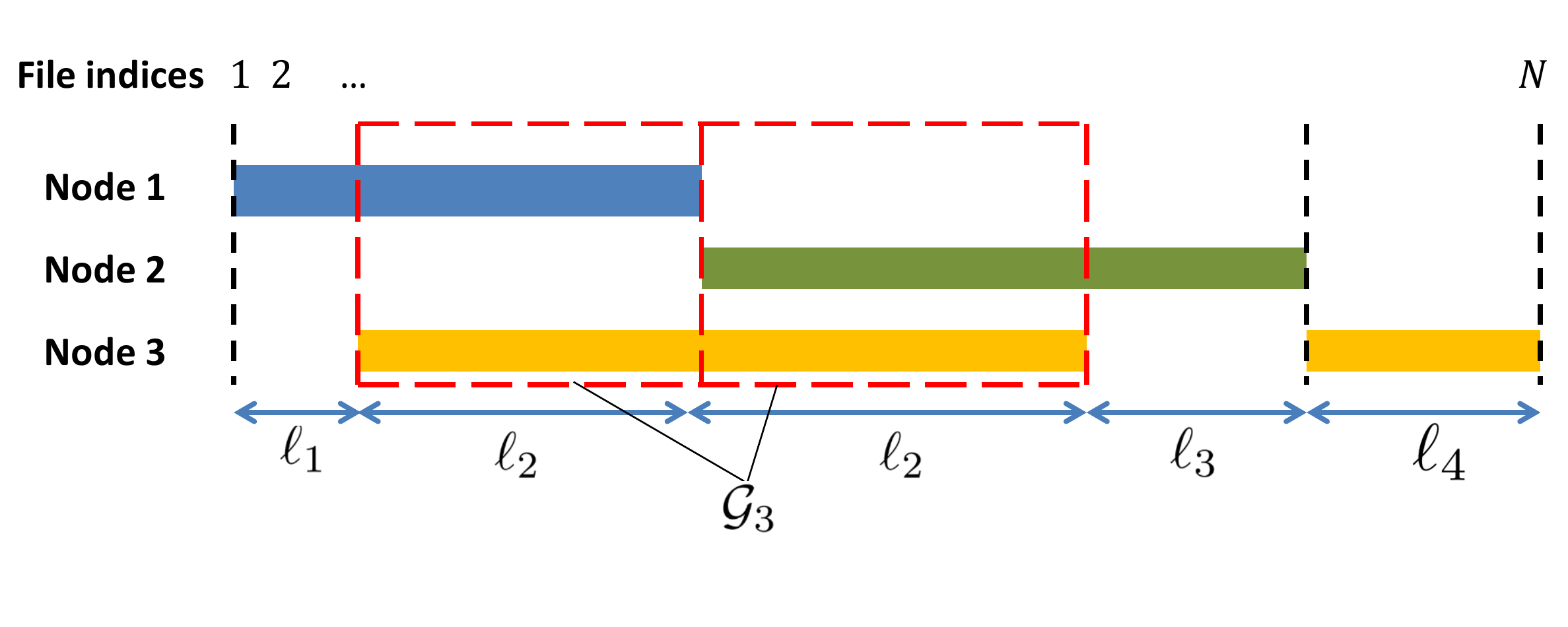}\vspace{-0.2in}
\caption{File placement for ${\mathcal R}_1$ where ${\ell}_1=M_1-(M-N)/2$, ${\ell}_2=(M-N)/2$, ${\ell}_3=M_2-(M-N)/2$, and ${\ell}_4=N-(M_1+M_2)$.}\label{caseR1}\vspace{-0.1in}
\end{figure}

We choose the file allocation $\mathcal{M}=(\mathcal{M}_1,\mathcal{M}_2,\mathcal{M}_3)$ as follows:
\begin{equation}
\begin{aligned}
\mathcal{M}_1&=[1:M_1],\\
\mathcal{M}_2&=[M_1+1:M_1+M_2],\\
\mathcal{M}_3&=[M_1+M_2+1:N]\cup[M_1-(M-N)/2+1:M_1+(M-N)/2].
\end{aligned}
\end{equation}
In order for the reader to understand more easily, we also illustrate the file allocation in Fig. \ref{caseR1}. With counting the length of the segment corresponding to each subset in Fig. \ref{caseR1}, the cardinality of each subset is given by:
\begin{eqnarray}
S_1&\!\!\!\!=\!\!\!\!&M_1-(M-N)/2,~~~S_{12} = 0,\notag\\
S_2&\!\!\!\!=\!\!\!\!&M_2-(M-N)/2,~~~S_{13} = (M-N)/2,\\
S_3&\!\!\!\!=\!\!\!\!&N-(M_1+M_2),~~~~~S_{23} = (M-N)/2.\notag
\end{eqnarray}
Hence, by applying the coding scheme specified in the proof of Lemma 1, we obtain:
\begin{equation}
{\mathcal L}=2(2N-M)+(M-N)/2=\frac{7}{2}N\hspace{.8mm}-\hspace{.8mm}\frac{3}{2}M.
\end{equation}

\subsubsection{$M_3>N+M_1-M_2$ (Regime $\mathcal{R}_4$)}

\begin{figure}[!t] \centering 
\includegraphics[trim = 0in 1.6in .5in 2in, clip,width=0.4\textwidth]{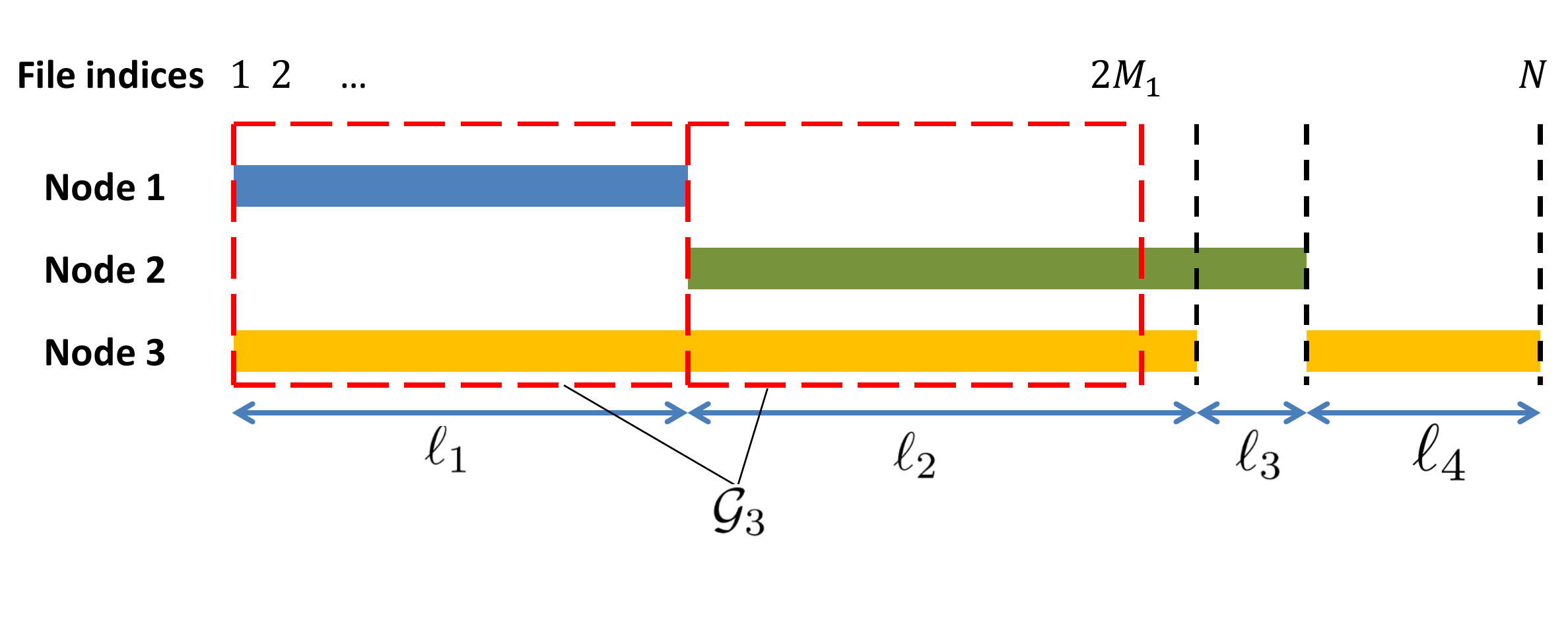}\vspace{-0.2in}
\caption{File placement for ${\mathcal R}_4$ where ${\ell}_1=M_1$, ${\ell}_2=M_2+M_3-N$, ${\ell}_3=N-M_3$, and ${\ell}_4=N-(M_1+M_2)$.}\label{caseR4}
\end{figure}

For this regime, we consider the file allocation $\mathcal{M}=(\mathcal{M}_1,\mathcal{M}_2,\mathcal{M}_3)$ as follows:
\begin{eqnarray}
\mathcal{M}_1&\!\!\!\!=\!\!\!\!&[1:M_1],\notag\\
\mathcal{M}_2&\!\!\!\!=\!\!\!\!&[M_1+1:M_1+M_2],\\
\mathcal{M}_3&\!\!\!\!=\!\!\!\!&[M_1+M_2+1:N]\cup [1:(M-N)].\notag
\end{eqnarray}
which are also shown in Fig. \ref{caseR4}, where the length of $L_1$, $L_2$, $L_3$ can be simply calculated via Lemma 1. Thus, the cardinality of each subset is given by:
\begin{eqnarray}
S_1&\!\!\!\!=\!\!\!\!&0,~~~~~~~~~~~~~~~~~~~~~~S_{12} = 0,\notag\\
S_2&\!\!\!\!=\!\!\!\!&N-M_3,~~~~~~~~~~~~~S_{13} = M_1,\notag\\
S_3&\!\!\!\!=\!\!\!\!&N-(M_1+M_2),~~~S_{23} = M_2+M_3-N.
\end{eqnarray}
Hence, based on Lemma 1, we have:
\begin{equation}
{\mathcal L}=2(2N\!-\!M)+(M_2\!+\!M_3\!-\!N)=3N-(M_1+M).
\end{equation}

\subsection{$M\leq 2N$ and $M_1+M_2> N$}


\subsubsection{$M_3\leq 3N-M_1-3M_2$ (Regime $\mathcal{R}_2$)}

\begin{figure}[!t] \centering 
\includegraphics[trim = 0in 1.7in .5in 1.5in, clip,width=0.4\textwidth]{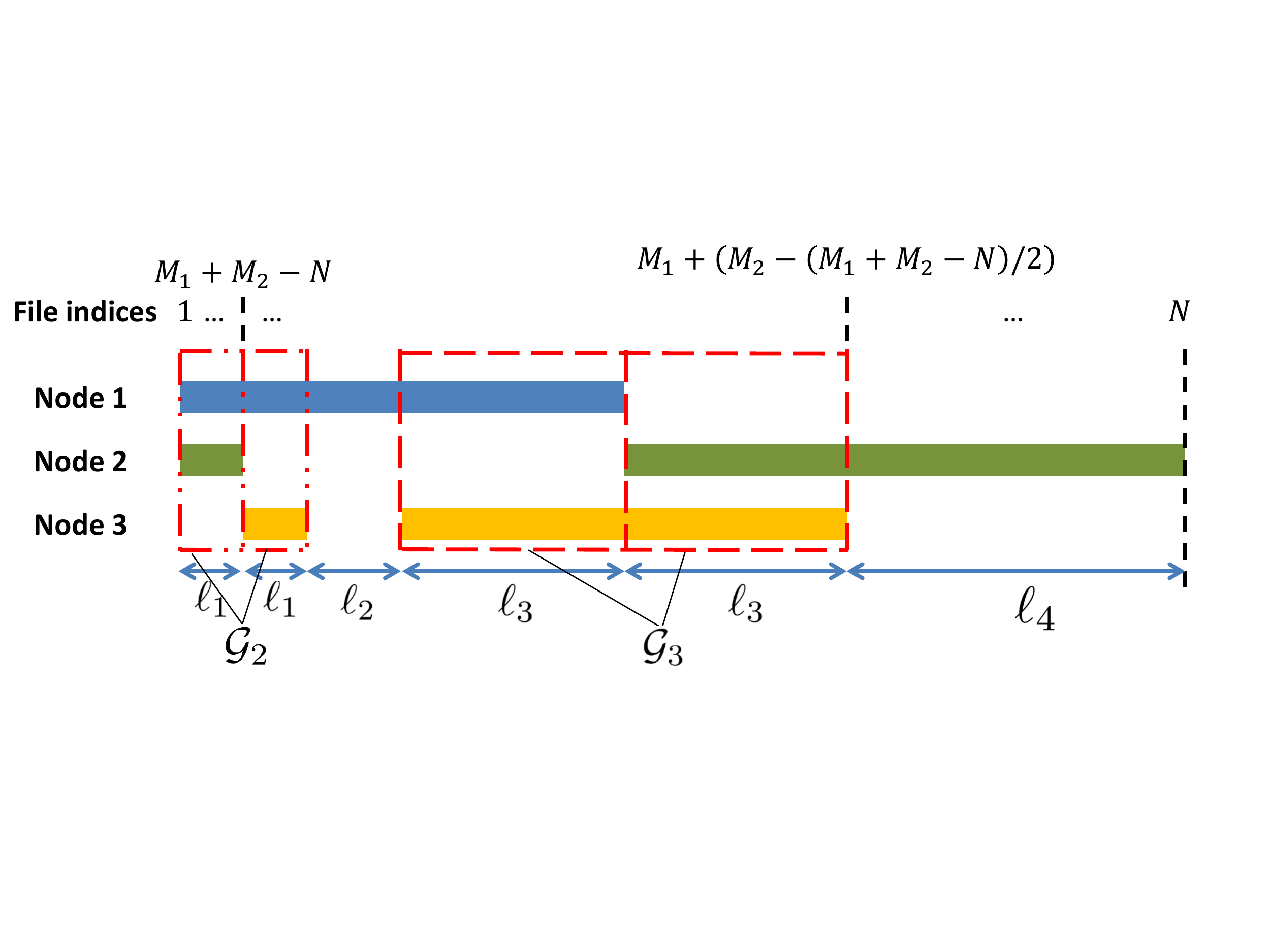}\vspace{-0.2in}
\caption{File placement for ${\mathcal R}_2$ where ${\ell}_1=M_1+M_2-N$, ${\ell}_2=M_1-2(M_1+M_2-N)-(M_3-(M_1+M_2-N))/2$, ${\ell}_3=(M_3-(M_1+M_2-N))/2$, and ${\ell}_4=N-M_1-(M_3-(M_1+M_2-N))/2$.}\label{caseR2}\vspace{-0.1in}
\end{figure}

We design the file allocation $\mathcal{M}=(\mathcal{M}_1,\mathcal{M}_2,\mathcal{M}_3)$, depicted in Fig. \ref{caseR2}, as follows:
\begin{eqnarray}
\mathcal{M}_1&\!\!\!\!=\!\!\!\!&[1:M_1],\notag\\
\mathcal{M}_2&\!\!\!\!=\!\!\!\!&[M_1+1:N]\cup [1:M_1+M_2-N],\\
\mathcal{M}_3&\!\!\!\!=\!\!\!\!&[M_1+M_2-N+1:2(M_1+M_2-N)]\cup \Big[M_1\!-\!\frac{M_3\!-\!(M_1\!+\!M_2\!-\!N)}{2}\!+\!1\!:\!M_1\!+\!\frac{M_3\!-\!(M_1\!+\!M_2\!-\!N)}{2}\Big]\notag
\end{eqnarray}
Thus, the cardinality of each subset is given by:
\begin{eqnarray}
\begin{array}{lll}
S_1=M_1\!-2(M_1\!+M_2\!-N)\!-(M_3\!-(M_1\!+M_2\!-N))/2,&&S_{12} = M_1+M_2-N,\\
S_2=N-M_1-(M_3-(M_1+M_2-N))/2,&&S_{13} = M_1\!+M_2\!-N\!+(M_3\!-(M_1\!+M_2\!-N))/2,\\
S_3=0,&&S_{23} = (M_3-(M_1+M_2-N))/2.
\end{array}
\end{eqnarray}
Hence, according to Lemma 1, we obtain:
\begin{eqnarray}
{\mathcal L}=2(2N-M)+M_1+M_2-N+\frac{M_3-(M_1+M_2-N)}{2}=\frac{7}{2}N-\frac{3}{2}M.
\end{eqnarray}

\subsubsection{$M_3>3N-M_1-3M_2$}\label{case:3Meq2N}


\begin{figure}[!b] \centering 
\includegraphics[trim = 0in 1.8in .5in 2in, clip,width=0.4\textwidth]{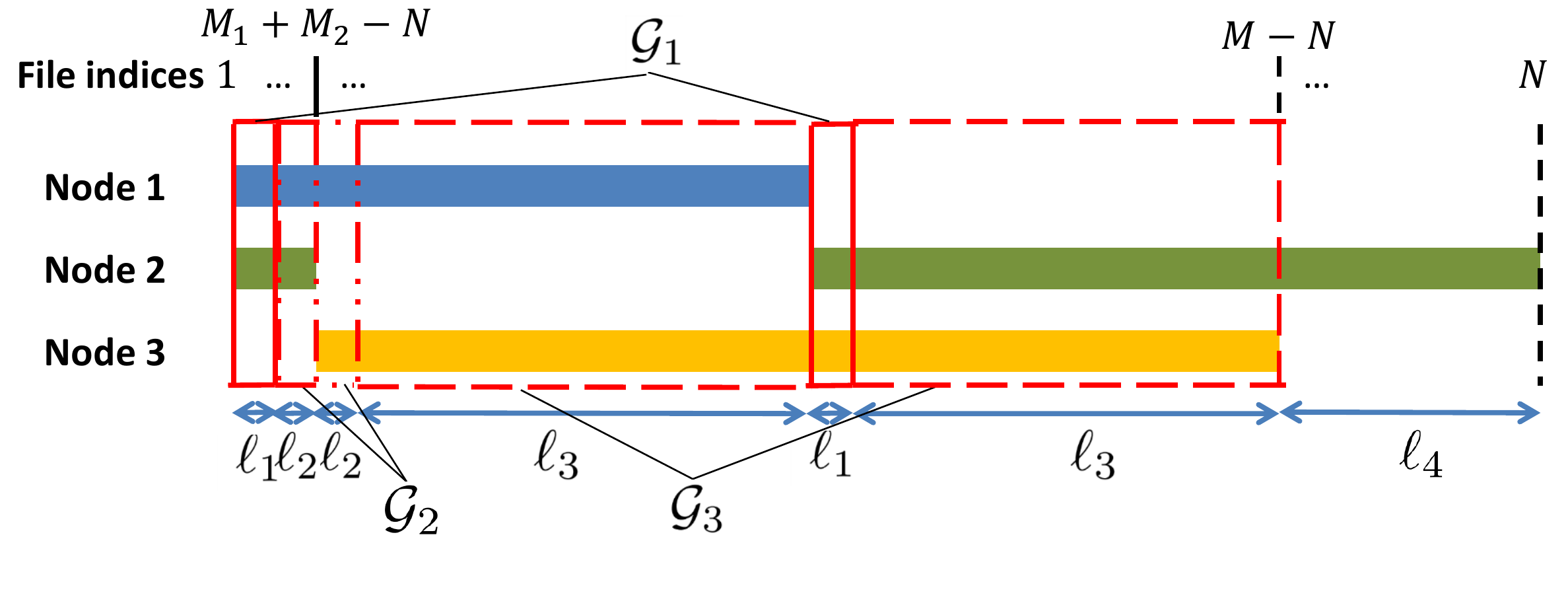}\vspace{-0.2in}
\caption{File placement for ${\mathcal R}_3$ where ${\ell}_1=\frac{M_1+M_2+M_3-N}{2}-(N-M_2)$, ${\ell}_2=\frac{M_1+M_2+M_3-N}{2}-(M_2+M_3-N)$, ${\ell}_3=M_2+M_3-N-L_1$, and ${\ell}_4=2N-M$.}\label{caseR3}
\end{figure}

\begin{figure}[!b] \centering 
\includegraphics[trim = 0in 1.7in .5in 2in, clip,width=0.4\textwidth]{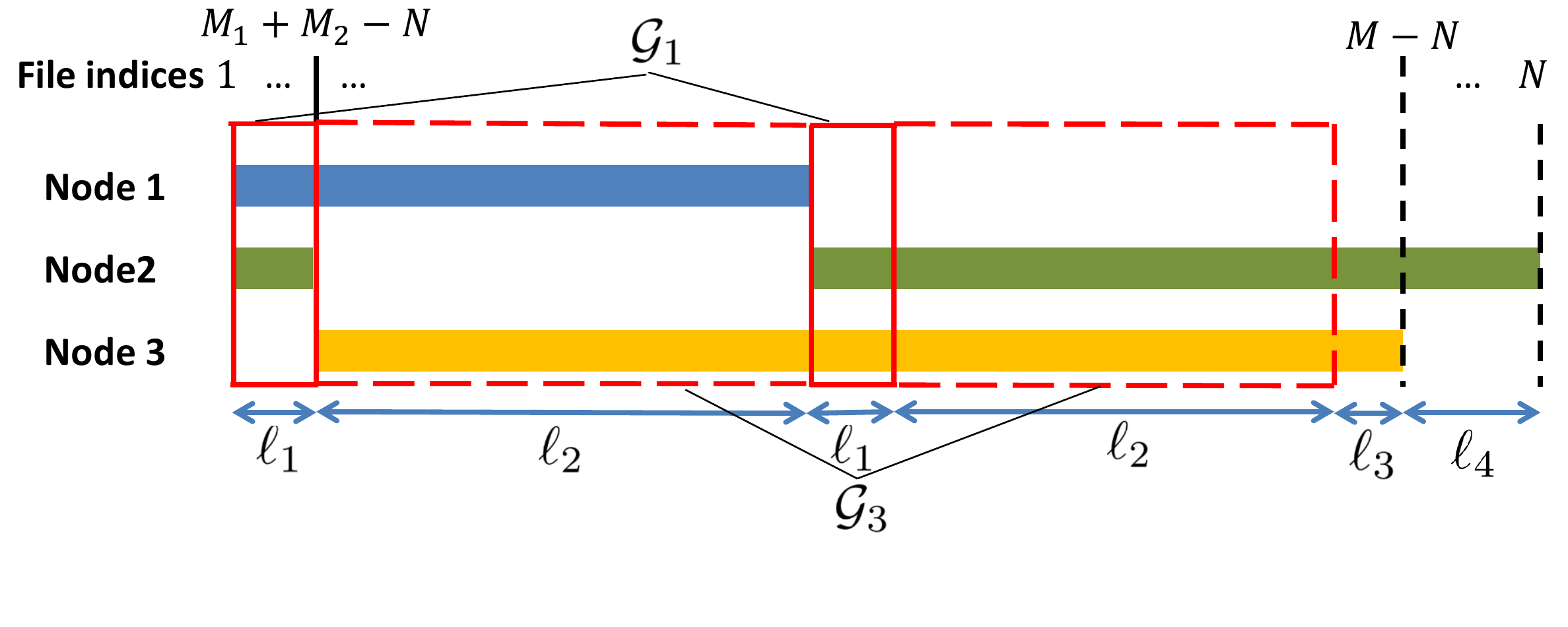}\vspace{-0.2in}
\caption{File placement for ${\mathcal R}_5$ where ${\ell}_1=M_1+M_2-N$, ${\ell}_2=N-M_2$, ${\ell}_3=M_2+M_3-N-M_1$, and ${\ell}_4=2N-M$.}\label{caseR5}\vspace{-0.1in}
\end{figure}

The file allocation $\mathcal{M}=(\mathcal{M}_1,\mathcal{M}_2,\mathcal{M}_3)$ is as follows:
\begin{eqnarray}
\mathcal{M}_1&\!\!=\!\!&[1:M_1],\notag\\
\mathcal{M}_2&\!\!=\!\!&[M_1+1:N]\cup [1:M_1+M_2-N],\\
\mathcal{M}_3&\!\!=\!\!&[M_1+M_2-N+1:M-N].\notag
\end{eqnarray}
Thus, the cardinality of each subset is given by:
\begin{eqnarray}
S_1&\!\!\!\!=\!\!\!\!&0,~~~~~~~~~~~~~~S_{12} = M_1+M_2-N,\notag\\
S_2&\!\!\!\!=\!\!\!\!&2N-M,~~~~~S_{13} = N-M_2,\\
S_3&\!\!\!\!=\!\!\!\!&0,~~~~~~~~~~~~~~S_{23} = M_2+M_3-N.\notag
\end{eqnarray}
According to Lemma 1, we have the results in the following two cases:

{\bf Case 1}: $M_3\leq  N+M_1-M_2$ (Regime $\mathcal{R}_3$)
\begin{eqnarray}
{\mathcal L}=2(2N-M)+(M-N)/2=\frac{7}{2}N-\frac{3}{2}M.
\end{eqnarray}

{\bf Case 2}: $M_3>N+M_1-M_2 $ (Regime $\mathcal{R}_5$)
\begin{eqnarray}
{\mathcal L}=2(2N\!-\!M)\!+\!M_2\!+\!M_3\!-\!N=3N-(M_1+M).
\end{eqnarray}

The corresponding file allocation for regimes $\mathcal{R}_3$ and $\mathcal{R}_5$ are shown in Fig. \ref{caseR3} and Fig. \ref{caseR5}, respectively.

\subsection{$M> 2N$}


\begin{figure}[!t] \centering 
\includegraphics[trim = 0in 1.8in .5in 2in, clip,width=0.40\textwidth]{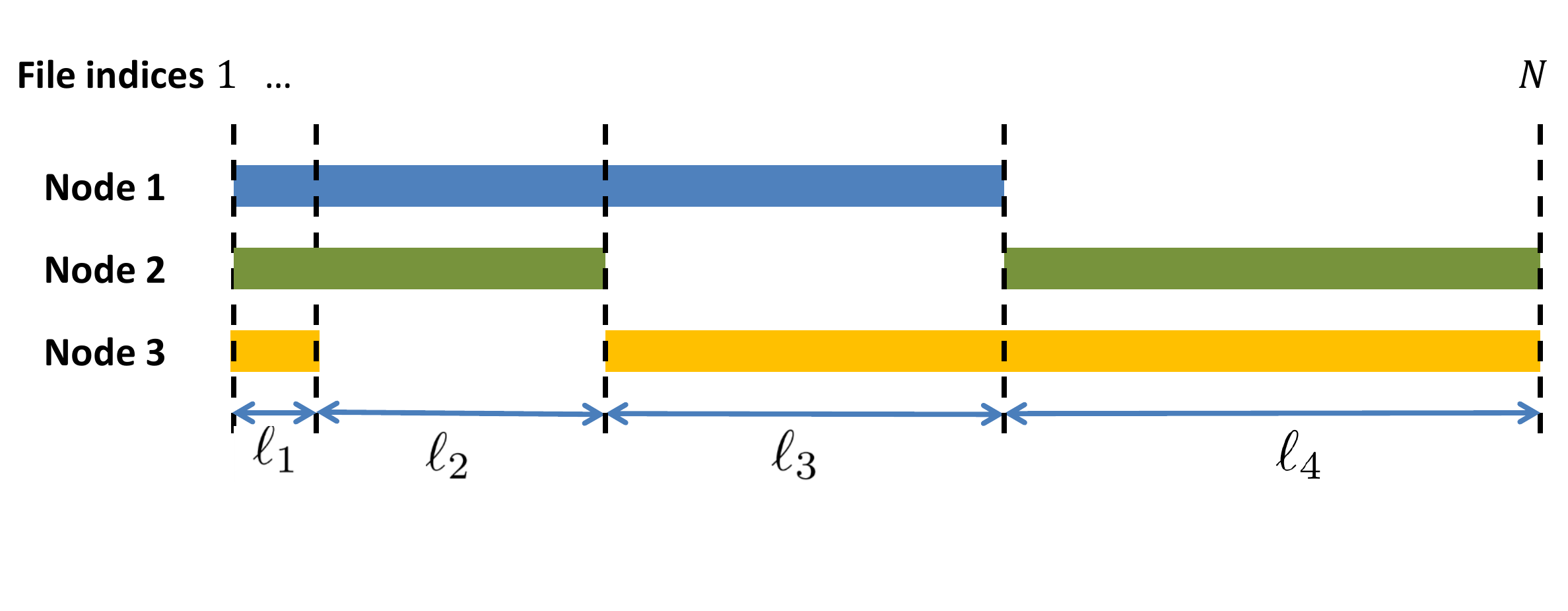}\vspace{-0.2in}
\caption{File placement for ${\mathcal R}_6$ where ${\ell}_1=M-2N$, ${\ell}_2=N-M_3$, ${\ell}_3=N-M_2$, and ${\ell}_4=N-M_1$.}\label{caseR6}\vspace{-0.1in}
\end{figure}

\begin{figure}[!t] \centering 
\includegraphics[trim = 0in 2in .5in 2in, clip,width=0.40\textwidth]{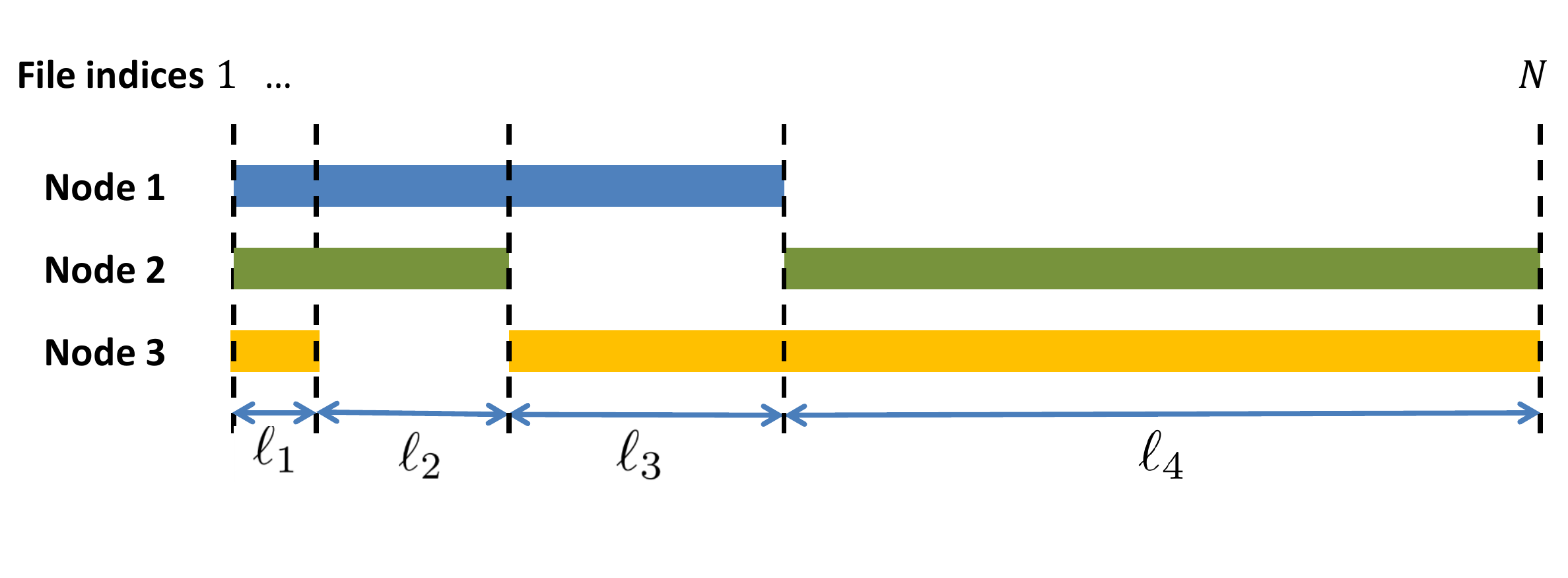}\vspace{-0.2in}
\caption{File placement for ${\mathcal R}_7$ where ${\ell}_1=M-2N$, ${\ell}_2=N-M_3$, ${\ell}_3=N-M_2$, and ${\ell}_4=N-M_1$.}\label{caseR7}\vspace{-0.1in}
\end{figure}

We consider the following file allocation $\mathcal{M}=(\mathcal{M}_1,\mathcal{M}_2,\mathcal{M}_3)$:
\begin{eqnarray}
\mathcal{M}_1&\!\!\!\!=\!\!\!\!&[1:M_1],\notag\\
\mathcal{M}_2&\!\!\!\!=\!\!\!\!&[M_1+1:N]\cup [1:M_1+M_2-N],\\
\mathcal{M}_3&\!\!\!\!=\!\!\!\!&[M_1+M_2-N+1:N]\cup [1:M_1+M_2+M_3-2N].\notag
\end{eqnarray}
Thus, the cardinality of each subset is given by:
\begin{eqnarray}
S_{123} = M-2N,~~~~S_1&\!\!\!\!=\!\!\!\!&0,~~~~S_{12} = M_1+M_2-N-(M-2N)=N-M_3,\notag\\
S_2&\!\!\!\!=\!\!\!\!&0,~~~~S_{13} = M_1+M_3-N-(M-2N)=N-M_2,\notag\\
S_3&\!\!\!\!=\!\!\!\!&0,~~~~S_{23} = M_2+M_3-N-(M-2N)=N-M_1.
\end{eqnarray}
It is straightforward to see that after removing the subset $\mathcal{S}_{123}$ from the sets $\mathcal{S}_{k},~k=1,2,3$, respectively, the remaining subsets $\mathcal{S}_{k}\setminus \mathcal{S}_{123},~k=1,2,3$ fall into the setting that we have identified in Section \ref{case:3Meq2N}). Specifically, we need to consider the two cases where $S_{23}\leq S_{12}+S_{13}$ and where $S_{23}> S_{12}+S_{13}$ due to $S_{12}\leq S_{13} \leq S_{23}$. Thus, according to Lemma 1, we have the results in the following two cases:

{\bf Case 1}: If $S_{23}\!\leq\! S_{12}\!+\!S_{13}$, then $ M_2\!+\!M_3\!-\!M_1\!\leq\! N$ (Regime $\mathcal{R}_6$)
\begin{eqnarray}
{\mathcal L}=\frac{S_{12}+S_{13}+S_{23}}{2}=\frac{3}{2}N-\frac{1}{2}M.
\end{eqnarray}

{\bf Case 2}: If $S_{23}\!>\! S_{12}\!+\!S_{13}$, then $ M_2\!+\!M_3\!-\!M_1\!>\! N$ (Regime $\mathcal{R}_7$)
\begin{eqnarray}
{\mathcal L}=S_{23}=N-M_1.
\end{eqnarray}

The file allocation corresponding to regimes $\mathcal{R}_6$ and $\mathcal{R}_7$ are shown in Fig.\ref{caseR6} and Fig.\ref{caseR7}, respectively.

\section{Converse of Theorem 1}\label{sec:converse}

Applying the result in \emph{Lemma 1} in \cite{Avestimehr_CDC}, letting $K=3$ and translating their notation into those we use, i.e., $1)$ converting the communication load defined in \cite{Avestimehr_CDC} into ours, $2)$ $a^1_{\mathcal{M}}=S_1+S_2+S_3$ and $3)$ $a^2_{\mathcal{M}}=S_{12}+S_{13}+S_{23}$ where $a^j_{\mathcal{M}}$ represents the number files stored at $j$ nodes only, we have the following corollary:

{\bf Corollary 1:} ${\mathcal L}_{\mathcal{M}}\geq 2(S_1+S_2+S_3)+\frac{1}{2}(S_{12}+S_{13}+S_{23})$.

\begin{remark}
It can be easily seen that the right-hand side of inequality above is the same as the right-hand side of (3) in Lemma 1 when the triangle inequality  $S_{12}+S_{13}+S_{23}-\max S_{ij}\geq \max S_{ij}$ is satisfied.
\end{remark}

Next, considering {\em any} possible file allocation ${\mathcal M}$ and {\em any} coding scheme, we will provide the lower bounds on ${\mathcal L}_{\mathcal M}$.

\subsection{Converse for ${\mathcal P}\in {\mathcal R}_1 \cup {\mathcal R}_2 \cup {\mathcal R}_3$: ${\mathcal L}_{\mathcal M}\geq \frac{7}{2}N-\frac{3}{2}M$}

According to our definition, we have:
\begin{eqnarray}
&&\left\{\begin{array}{l}S_1+S_{12}+S_{13}+S_{123}=M_1\\S_2+S_{12}+S_{23}+S_{123}=M_2\\S_3+S_{13}+S_{23}+S_{123}=M_3\end{array}\right.
\end{eqnarray}
which leads to
\begin{eqnarray}
S_{12}+S_{13}+S_{23}=\frac{(\sum_{k=1}^3{M_k})-(\sum_{i=1}^3{S_i})-3S_{123}}{2}.\label{eqn:S1_S2_S3}
\end{eqnarray}
Due to $S_1+S_2+S_3+S_{12}+S_{13}+S_{23}+S_{123}=N$, we have
\begin{eqnarray}
&&S_{123}=M+\Big(\sum_{i=1}^3{S_i}\Big)-2N.\label{eqn:S123}
\end{eqnarray}
Substituting (\ref{eqn:S123}) into (\ref{eqn:S1_S2_S3}) and then expressing $S_{12}+S_{13}+S_{23}$ with $S_{1}+S_{2}+S_{3}$, we can rewrite Corollary 1 as follows:
\begin{eqnarray}
{\mathcal L}_{\mathcal M}\geq \frac{3}{2}N-\frac{1}{2}M+\sum_{i=1}^3{S_i}.\label{eqn:TM}
\end{eqnarray}
Since $M\leq 2N$, we must have $S_1\!+\!S_2\!+\!S_3\geq 2N\!-\!M$ for every possible file allocation. Thus, we obtain the bound:
\begin{eqnarray}
{\mathcal L}_{\mathcal M}\geq \frac{3}{2}N-\frac{1}{2}M+\sum_{k=1}^3{S_k}\geq \frac{7}{2}N-\frac{3}{2}M.
\end{eqnarray}

\subsection{Converse for ${\mathcal P}\in{\mathcal R}_6$: ${\mathcal L}_{\mathcal M}\geq \frac{3}{2}N-\frac{1}{2}M$}


Following from (\ref{eqn:TM}), since we always have $S_1+S_2+S_3\geq 0$, we can directly obtain:
\begin{eqnarray}
{\mathcal L}_{\mathcal M}\geq \frac{3}{2}N-\frac{1}{2}M+\sum_{k=1}^3 S_k\geq \frac{3}{2}N-\frac{1}{2}M.
\end{eqnarray}

Observations of Theorem 1 reveal that the two bounds shown above apply to $\mathcal{R}_1$, $\mathcal{R}_2$, $\mathcal{R}_3$ and $\mathcal{R}_6$. What remains to be shown is the converse for $\mathcal{R}_4$, $\mathcal{R}_5$ and $\mathcal{R}_7$, respectively. Next, We provide two new propositions in the following.

\subsection{Converse for ${\mathcal P}\in{\mathcal R}_7$: ${\mathcal L}_{\mathcal M}\geq N-M_1$}


This bound is essentially a ``cut-set" bound, because intuitively server 1 needs at least a total of $N-M_1$ equations to collect its desired intermediate symbols. The formal proof is briefly written as follows:
\begin{eqnarray}
{\mathcal L}_{\mathcal M}&\!\!\!\!\geq \!\!\!\!& H(X_1,X_2,X_3) \\
&\!\!\!\!\geq \!\!\!\!&  H(X_2,X_3|V_{:,\mathcal{M}_1})\\
&\!\!\!\!\geq \!\!\!\!& I(X_2,X_3;V_{1,:}|V_{:,\mathcal{M}_1})\label{eqn:bound3_decode}\\
&\!\!\!\!=\!\!\!\!& H(V_{1,:}|V_{:,\mathcal{M}_1})-\underbrace{H(V_{1,:}|X_2,X_3,V_{:,\mathcal{M}_1})}_{\overset{(a)}{=}0}\ \ \ \\
&\!\!\!\!=\!\!\!\!&N-M_1
\end{eqnarray}
where $(a)$ is due to the decoding requirement at node 1.

\subsection{Converse for ${\mathcal P}\in{\mathcal R}_4\cup{\mathcal R}_5$: ${\mathcal L}_{\mathcal M}\geq 3N-(M+M_1)$}


Intuitively, this bound $\mathcal{L}_{\mathcal{M}}\geq (N-M_1)+(2N-M)$ can be interpreted as number of equations needed to meet the ``cut-set" bound shown above and associated with the files stored at only one node and sent to another node. 
We first provide two inequalities in the following two lemmas.

{\bf Lemma 2:} $H(X_1|X_2,X_3)\geq 2S_1$.

{\it Proof:} The proof is readily shown as follows:
\begin{eqnarray}
H(X_1|X_2,X_3)&\!\!\!\!\geq\!\!\!\!& H(X_1|X_2,X_3,V_{:,\mathcal{M}_2},V_{:,\mathcal{M}_3})\label{eqn:bound4_decode2}\\
&\!\!\!\!\geq\!\!\!\!& I(X_1;V_{2,:},V_{3,:}|X_2,X_3,V_{:,\mathcal{M}_2},V_{:,\mathcal{M}_3})\\
&\!\!\!\!=\!\!\!\!& \underbrace{H(V_{2,:}|X_2,X_3,V_{:,\mathcal{M}_2},V_{:,\mathcal{M}_3})}_{\geq S_1}+\underbrace{H(V_{3,:}|V_{2,:},X_2,X_3,V_{:,\mathcal{M}_2},V_{:,\mathcal{M}_3})}_{\geq S_1}\notag\\
&\!\!\!\!\!\!\!\!&- \underbrace{H(V_{2,:},V_{3,:}|X_1,X_2,X_3,V_{:,\mathcal{M}_2},V_{:,\mathcal{M}_3})}_{\overset{(a)}{=}0}\ \ \ \\
&\!\!\!\!\geq\!\!\!\!& 2S_1
\end{eqnarray}
where $(a)$ is due to the decoding requirements at node 2, 3.

\begin{remark}
This lemma essentially implies that the message $X_1$ must contain $S_1$ equations required by node 2 and another $S_1$ equations required by node 3, because node 2 and node 3 do not have any intermediate signals accosted with the files in $\mathcal{N}\setminus (\mathcal{M}_2 \cup \mathcal{M}_3)=\mathcal{S}_1$.
\end{remark}

{\bf Lemma 3:} $H(X_2,X_3|V_{1,:},V_{:,\mathcal{M}_1})\geq S_2+S_3$.

{\it Proof:} It can be proved by following the similar approach above and omitted due to the space limitation.

Finally, we derive the desired bound as follows.
\begin{eqnarray}
{\mathcal L}_{\mathcal M}&\!\!\!\!\geq\!\!\!\!& H(X_1,X_2,X_3) \\
&\!\!\!\!=\!\!\!\!& H(X_2,X_3)+H(X_1|X_2,X_3)\\
&\!\!\!\!\geq\!\!\!\!& H(X_2,X_3|V_{:,\mathcal{M}_1})+H(X_1|X_2,X_3)\\
&\!\!\!\!=\!\!\!\!& H(X_2,X_3|V_{:,\mathcal{M}_1}\!)\!+\!H(V_{1,:}|X_2,X_3,\!V_{:,\mathcal{M}_1}\!)\!+\!H(X_1|X_2,X_3\!)\ \ \ \ \ \\
&\!\!\!\!=\!\!\!\!& H(V_{1,:}|V_{:,\mathcal{M}_1})\!+\!H(X_2,X_3|V_{1,:},V_{:,\mathcal{M}_1})\!+\!H(X_1|X_2,X_3)\ \ \ \ \\
&\!\!\!\!\geq\!\!\!\!& N-M_1+H(X_2,X_3|V_{1,:},V_{:,\mathcal{M}_1})+H(X_1|X_2,X_3)\\
&\!\!\!\!\geq\!\!\!\!& N-M_1+S_2+S_3+2S_1\label{eqn:bound4_lemma12}\\
&\!\!\!\!\geq\!\!\!\!& N-M_1+2N-(M_1+M_2+M_3)+S_1\\
&\!\!\!\!\geq\!\!\!\!& N-M_1+2N-M
\end{eqnarray}
where (\ref{eqn:bound4_lemma12}) is obtained due to Lemma 2 and Lemma 3.

The union of the inequalities derived above cover all the regimes specified in Theorem 1. This completes the entire converse proof of Theorem 1. In addition, observations of the 4 inequalities reveal that each inequality is a valid lower bound in every regime, but they are not simultaneously active.

\section{Algorithm of the Achievability for the General $K$ Server Nodes}\label{sec:generalK}

As we introduced earlier, due to the difficulties of seeking information theoretic result for $K>3$, we provide an algorithm to investigate the achievability of the communication load.

Before presenting the algorithm, we first summarize the main idea behind our algorithm. Specifically, given a file allocation $\mathcal{M}=\{\mathcal{M}_k\}_{k=1}^K$, we can identify a total of $2^K-1$ subsets (similar to 7 subsets defined for the case $K=3$) by considering the relationship among the $K$ subsets $\mathcal{M}_1,\cdots,\mathcal{M}_K$. Thus, we decompose the system into up to $K$ parallel subsystems where in the $j^{th}$ subsystem for each $j\in\mathcal{K}$, every file is stored at $j$ nodes only. Then for each subsystem, we will find the coding opportunities as many as possible, by using coding schemes developed for $K=3$ in Section \ref{sec:achievability} in this paper and in \cite{Avestimehr_CDC}. Since each coding equation involves $j$ intermediate values, we can save at least $j-1$ transmissions compared to the communication load $\mathcal{L}^j_{\textrm{u}}$ of the uncoded scheme. Hence, the communication load contributed by the $j^{th}$ subsystem is given by $\mathcal{L}^j=\mathcal{L}^j_{\textrm{u}}-(j-1)l_j$ where $l_j$ is the number of equations we can create. Next, we obtain $\mathcal{L}=\sum_{j\in\mathcal{K}} \mathcal{L}^j$, i.e., adding up the communication load contributed by each subsystem. Since the explicit expressions of the $2^K-1$ subsets depend on the choice of file allocation $\mathcal{M}$, we treat $\mathcal{M}$ as undetermined variables and choose the total communication load $\mathcal{L}$ as the objective function which is a linear function of the variables we define. Finally, resolving the linear programming problem we formulate, we are able to obtain a feasible solution.

Since our proposed algorithm might not be easy to read, in the following we will first provide two specific examples for $K=3$ and $K=4$ before presenting the algorithm for the systems with general $K>4$.

\subsection{Example 1: $K=3$}

Let us start with the simplest setting with $K=3$. Although this setting has already been completely resolved in the prior sections, we will readily show an alternative approach of developing an algorithm to formulate the original problem into an optimization problem, while getting rid of the 7 regime classification. Given the setting with $(M_1,M_2,M_3,N)$, the algorithm, which results in the same conclusion as in Theorem 1, is shown in the following.

\begin{itemize}
\item{Step 0:} Initialize the communication load $\mathcal{L}=0$, and the set of constraints $\mathcal{E}=\emptyset$.

\item{Step 1:} Consider the subsystem $j=1$ where we only have the files in the 3 subsets $\mathcal{S}_1$, $\mathcal{S}_2$, $\mathcal{S}_3$. According to Section \ref{sec:achievability}, we can directly obtain the communication load function:
    \begin{eqnarray}
    \mathcal{L}^1=2(S_1+S_2+S_3).
    \end{eqnarray}

\item{Step 2:} Consider the subsystem $j=K-1=2$ where we only have $\mathcal{S}_{12}$, $\mathcal{S}_{13}$, $\mathcal{S}_{23}$. Recall that in Lemma 1 in Section \ref{sec:achievability}, we develop the function $g(\cdot)$ based on the subsets where each file is stored at $K-1=2$ nodes only. Denoting the number of files in $\mathcal{S}_{12}$, $\mathcal{S}_{13}$, $\mathcal{S}_{23}$ as undetermined non-negative variables $x_{12}$, $x_{13}$, $x_{23}$, we can obtain the following inequalities:
    \begin{eqnarray}
    x_{21}+x_{22}\leq S_{12},\notag\\
    x_{21}+x_{23}\leq S_{13},\\
    x_{22}+x_{23}\leq S_{23}.\notag
    \end{eqnarray}
    In addition, we also have $x_{2q}\geq 0$, for $q=1,2,3$, and $S_{12}\geq 0,S_{13}\geq 0,S_{23}\geq 0$. Add all these $9$ inequalities to $\mathcal{E}$.

\item{Step 3:} Then the communication load function $\mathcal{L}^2$ can be written as follows:
\begin{eqnarray}
\mathcal{L}^2=(S_{12}+S_{13}+S_{23})-(x_{21}+x_{22}+x_{23}).
\end{eqnarray}

\item{Step 4:} Consider the subsystem $j=K=3$ where we only have $\mathcal{S}_{123}$. Clearly, $\mathcal{L}^3=0$.

\item{Step 5:} Next, we obtain the entire communication load function $\mathcal{L}=\mathcal{L}^1+\mathcal{L}^2+\mathcal{L}^3$. Meanwhile, we also have the following constraints over the entire files:
\begin{eqnarray}
S_1+S_2+S_3+S_{12}+S_{13}+S_{23}+S_{123}&\!\!\!\!=\!\!\!\!&N,\\
S_1+S_2+S_3+2(S_{12}+S_{13}+S_{23})+3S_{123}&\!\!\!\!=\!\!\!\!&M_1+M_2+M_3,
\end{eqnarray}
and the following constraints over the files allocated at each server:
\begin{eqnarray}
S_1+S_{12}+S_{13}+S_{123}&\!\!\!\!=\!\!\!\!&M_1,\notag\\
S_2+S_{12}+S_{23}+S_{123}&\!\!\!\!=\!\!\!\!&M_2,\\
S_3+S_{13}+S_{23}+S_{123}&\!\!\!\!=\!\!\!\!&M_3.\notag
\end{eqnarray}
Add all these $K+2=5$ equations to $\mathcal{E}$.

\item{Step 6:} Finally, after removing redundant variables and constraints, we translate seeking the communication load problem into resolving the linear optimization problem:
\begin{eqnarray*}
\textrm{min} && \mathcal{L}=2(S_1+S_2+S_3)+(S_{12}+S_{13}+S_{23})-(x_{21}+x_{22}+x_{23}) \\
\textrm{subject to} && \mathcal{E}\triangleq \left\{\begin{array}{l}
x_{21}+x_{22}\leq S_{12},\\
x_{21}+x_{23}\leq S_{13},\\
x_{22}+x_{23}\leq S_{23},\\
x_{2q}\geq 0,~~~q=1,2,3.\\
S_1+S_{12}+S_{13}+S_{123}=M_1,\\
S_2+S_{12}+S_{23}+S_{123}=M_2,\\
S_3+S_{13}+S_{23}+S_{123}=M_3,\\
S_1+S_2+S_3+S_{12}+S_{13}+S_{23}+S_{123}=N,\\
S_k\geq 0,~~~k=1,2,3,\\
S_{ij}\geq 0,~~~i,j=1,2,3,~j\neq j,\\
S_{123}\geq 0.
\end{array}\right.
\end{eqnarray*}
This linear optimization problem above can be easily resolved via several algorithms and programming tools.

\item{Step 7:} According to optimal solution we obtained above, we can readily determine file allocation $\mathcal{M}_k^o$ greedily for each server $k$ sequentially.

\end{itemize}

\begin{remark}
It can be readily seen that the linear optimization problem above is equivalent to the original problem, but we do not need to specify the boundaries between those 7 regimes associated with $M_1$, $M_2$, $M_3$ and $N$.
\end{remark}

\subsection{Example 2: $K=4$}

Considering the setting of $(\{M_k\}_{k=1}^4,N)$, the simplest setting beyond $K=3$, we can use the following algorithm to find an achievability.

\begin{itemize}
\item{Step 0:} Initialize the communication load $\mathcal{L}=0$, and the set of constraints $\mathcal{E}=\emptyset$.

\item{Step 1:} Consider the subsystem $j=1$ where we only have the files in the 4 subsets $\mathcal{S}_1$, $\mathcal{S}_2$, $\mathcal{S}_3$, $\mathcal{S}_4$. Similar to the subsystem $j=1$ for $K=3$ explained in Section \ref{sec:achievability}, we can directly obtain the communication load function:
    \begin{eqnarray}
    \mathcal{L}^1=3(S_1+S_2+S_3+S_4).
    \end{eqnarray}

\item{Step 2:} Then we consider the subsystem with $j=2$ where we only have $\{\mathcal{S}_{ij}\}_{(i,j)\in\mathcal{C}}$ and the set $\mathcal{C}$ is given by $\mathcal{C}=\{(1,2),(1,3),(1,4),(2,3),(2,4),(3,4)\}$. Considering the optimal coding scheme identified in \cite{Avestimehr_CDC} for the homogeneous system with $j=2$ and $K=4$, we have the following three file allocation methods to achieve the minimum communication load, i.e., $\{(1,2),(1,3),(2,4),(3,4)\}$, $\{(1,2),(1,4),(2,3),(3,4)\}$, and $\{(1,3),(1,4),(2,3),(2,4)\}$. Specifically, regarding the first set, it implies that if we allocate $L$ files to $K=4$ users, we will allocate the first quarter to node 1, 2, the second quarter to node 1, 3, the third quarter to node 2, 4, and the last quarter to node 3, 4.

\item{Step 3:} Denote the number of files for encoding by using the three methods above by 3 non-negative variables $x_{21}$, $x_{22}$, $x_{23}$. Clearly, we must have
    \begin{eqnarray}
    x_{21}+x_{22}\leq S_{12},~~~~x_{21}+x_{22}\leq S_{34},\notag\\
    x_{21}+x_{23}\leq S_{13},~~~~x_{21}+x_{23}\leq S_{24},\\
    x_{22}+x_{23}\leq S_{14},~~~~x_{22}+x_{23}\leq S_{23},\notag
    \end{eqnarray}
    due to the cardinality constraints. In addition, we add all these inequalities into $\mathcal{E}$. Also, the communication load $\mathcal{L}^2$ can be written as follows:
\begin{eqnarray}
\mathcal{L}^2=(K-j)(S_{12}+S_{13}+S_{14}+S_{23}+S_{24}+S_{34})-(j-1)K(x_{21}+x_{22}+x_{23})
\end{eqnarray}
where $K=4$ and $j=2$.

\item{Step 4:} Consider the subsystem $j=K-1=3$ where we only have $\mathcal{S}_{123}$, $\mathcal{S}_{124}$, $\mathcal{S}_{134}$, $\mathcal{S}_{234}$. Recall that in Lemma 1 for $K=3$ in Section \ref{sec:achievability}, we develop the function $g(\cdot)$ based on the subsets where each file is stored at $K-1=2$ nodes only. In fact, we generalize the function $g(\cdot)$ to achieve the information-theoretically optimal (i.e., minimum) communication load for $K=3$ to $K>3$. In particular, denote the number of files in $\mathcal{S}_{123}$, $\mathcal{S}_{124}$, $\mathcal{S}_{134}$ as undetermined non-negative variables $x_{31}$, $x_{32}$, $x_{33}$, $x_{34}$. Extending the main idea behind the proof of Lemma 1, we can obtain the following inequalities:
\begin{eqnarray}
x_{31}+x_{32}+x_{33}\leq S_{123},\notag\\
x_{31}+x_{32}+x_{34}\leq S_{124},\\
x_{31}+x_{33}+x_{34}\leq S_{134},\notag\\
x_{32}+x_{33}+x_{34}\leq S_{234}.\notag
\end{eqnarray}
In addition, we also have $x_{3q}\geq 0$, for $q=1,2,3,4$, and $S_{ijk}\geq 0$. Add all these $12$ inequalities to $\mathcal{E}$.

\item{Step 5:} Then the communication load function $\mathcal{L}^3$ can be written as follows:
\begin{eqnarray*}
\mathcal{L}^3\!=\!(S_{123}\!+S_{124}\!+S_{134}\!+S_{234})\!-\!2(x_{31}\!+x_{32}\!+x_{33}\!+x_{34}).
\end{eqnarray*}

\item{Step 6:} Consider the subsystem $j=K=4$ where we only have $\mathcal{S}_{1234}$. Clearly, $\mathcal{L}^4=0$.

\item{Step 7:} Next, we obtain the entire communication load function $\mathcal{L}=\sum_{j=1}^4\mathcal{L}^j$. Meanwhile, we also have the following constraints over the entire files:
\begin{eqnarray}
\sum_kS_k+\sum_{ij}S_{ij}+\sum_{ijk}S_{ijk}+S_{1234}&\!\!\!\!=\!\!\!\!&N,\notag\\
\sum_kS_k+2\sum_{ij}S_{ij}+3\sum_{ijk}S_{ijk}+S_{1234}&\!\!\!\!=\!\!\!\!&M,
\end{eqnarray}
and the constraints over the files allocated at each node:
\begin{eqnarray}
S_1\!+S_{12}\!+S_{13}\!+S_{14}\!+S_{123}\!+S_{124}\!+S_{134}\!+S_{1234}\!\!&\!\!\!\!=\!\!\!\!&\!\!M_1,\notag\\
S_2\!+S_{12}\!+S_{23}\!+S_{24}\!+S_{123}\!+S_{124}\!+S_{234}\!+S_{1234}\!\!&\!\!\!\!=\!\!\!\!&\!\!M_2,\ \ \ \ \ \ \\
S_3\!+S_{13}\!+S_{23}\!+S_{34}\!+S_{123}\!+S_{134}\!+S_{234}\!+S_{1234}\!\!&\!\!\!\!=\!\!\!\!&\!\!M_3,\notag\\
S_4\!+S_{14}\!+S_{24}\!+S_{34}\!+S_{124}\!+S_{134}\!+S_{234}\!+S_{1234}\!\!&\!\!\!\!=\!\!\!\!&\!\!M_4.\notag
\end{eqnarray}

Add all these $K+2=6$ equations to $\mathcal{E}$.

\item{Step 8:} Finally, after removing redundant variables and constraints, we translate seeking the communication load problem into resolving the linear optimization problem:
\begin{eqnarray*}
\textrm{min} ~~ \mathcal{L},~~~~~~~~\textrm{subject to} ~~ \mathcal{E}.
\end{eqnarray*}

\item{Step 9:} According to optimal solution we obtained above, we can determine file allocation $\{\mathcal{M}_k^o\}_{k\in\mathcal{K}}$ greedily for each node $k$ sequentially.

\end{itemize}



\begin{remark}
Based on what we have obtained so far, several interesting observations can be summarized as follows.
\begin{enumerate}

\item It can be easily seen that we consider the coding opportunities among the nodes' files in every $j^{th}$ subsystem, individually. Since we do not consider the coding opportunities across subsystems, this algorithm is suboptimal in general.

\item Observations of steps $2-3$ (for the subsystems with $1<j<K-1$) and steps $4-5$ (for the subsystem with $j=K-1$) reveal that they have the similar form, but their cost functions appear to be different. This is because for the subsystem with $j=K-1$, we generalize the coding idea that we identified for $K=3$ to the general setting with $K>3$, which further means that for the subsystem with $j=K-1$, our achievability is also information-theoretically optimal.

\item Recall that in Lemma 1 in Section \ref{sec:achievability}, the communication load includes the function of $g(\cdot)$, which is piece-wise linear. In contrast, still regarding the subsystem with $j=K-1$, $\mathcal{L}^3$ in Step 5 is a linear function. This is the key to formulate finding the communication load into resolving a linear programming problem.

\end{enumerate}

\end{remark}

\subsection{The Algorithm for General $K$}

Finally, we are going to show the algorithm for general $K$. Considering the setting of $(\{M_k\}_{k=1}^K,N)$ where $M_k>0$ for $\forall k\in \mathcal{K}$ and $\sum_{k=1}^K M_k\geq N$, we state the algorithm as follows.

\begin{itemize}
\item{Step 0:} (Initialization) Set $\mathcal{L}=0$, $\mathcal{E}=\emptyset$, $j=1$.

\item{Step 1:} Find out the collection $\mathcal{C}_j$ of every possible subsets in $\mathcal{K}$ with cardinality $j$: $\mathcal{C}_j\triangleq \{\mathcal{K'}~|~\mathcal{K'}\subset \mathcal{K},~|\mathcal{K'}|=j\}$, and its cardinality is given by $|\mathcal{C}_j|=\big(\!\!\begin{tiny}\begin{array}{c}K\\j\end{array}\end{tiny}\!\!\big)\triangleq P$. Note that each element in $\mathcal{C}_j$ is a subset or a tuple with $j$ indices.

\item{Step 2:} Find out the collection $\mathcal{C'}_j$ of every possible subsets in $\mathcal{C}_j$ with cardinality $K$ where the node indices from 1 to $K$ appear exactly $j$ times: $\mathcal{C'}_j\triangleq \{\mathcal{C}~|~\forall \mathcal{C}\in \mathcal{C}_j,~\sum {\bf 1}(k\in \mathcal{C})=j,~\textrm{for}~\forall k\in\mathcal{K}\}$. We denote its cardinality by $|\mathcal{C'}_j|=Q$.

\item{Step 3:} Claim $Q$ non-negative variables $x_{j1},x_{j2},\cdots,x_{jQ}$. Set $p=1$.

\item{Step 4:} Add up all the variables $\{x_{jq}\}$ if the $q^{th}$ element in $\mathcal{C'}_j$ contains $(\mathcal{C}_j)_p$, the $p^{th}$ subset or tuple in $\mathcal{C}_j$ for $q=1,\cdots,Q$, and upper bound it by a non-negative $S_{(\mathcal{C}_j)_p}$. That is,
\begin{eqnarray*}
\sum_{q=1}^Q x_{jq}{\bf 1}((\mathcal{C}_j)_p\in(\mathcal{C'}_j)_q)\leq S_{(\mathcal{C}_j)_p}.
\end{eqnarray*}
In addition, note that $x_{jq}\geq 0$, for $q=1,2,\cdots,Q$, and $S_{(\mathcal{C}_j)_p}\geq 0$. Add all these $Q+2$ inequalities to $\mathcal{E}$.

\item{Step 5:} If $p<Q$, then $p=p+1$ and go back to Step 4; Otherwise, go to Step 6.

\item{Step 6:} Develop the cost function $\mathcal{L}^j$ as follows. Without coding, we need a total of $(K-j)\sum_{p=1}^P S_{(\mathcal{C}_j)_p}$ transmissions. Note that we create a total of $\sum_{q=1}^Q x_{jq}$ collections, each with $K$ subsets and each file exactly mapped to $j$ servers. After extending  the encoding scheme described in \cite{Avestimehr_CDC} to the homogeneous setting, we can save $(K-j)(1-\frac{1}{j})Kx_{jq}$ for the $q^{th}$ collection. Thus, we save a total of $(K-j)(1-\frac{1}{j})(\sum_{q=1}^Q x_{jq})$ transmissions. Hence, the cost function is given by:
\begin{eqnarray*}
\mathcal{L}^j=(K-j)\sum_{p=1}^P S_{(\mathcal{C}_j)_p}-K(K-j)\bigg(1-\frac{1}{j}\bigg)\bigg(\sum_{q=1}^Q x_{jq}\bigg).
\end{eqnarray*}
Then we update the objective function as $\mathcal{L}=\mathcal{L}+\mathcal{L}^j$.

\item{Step 7:} Let $j=j+1$. If $j<K-1$, then go back to Step 1; Otherwise, go to Step 8.

\item{Step 8:} Find out the collection $\mathcal{C}_{K-1}$ of every possible subsets in $\mathcal{K}$ with cardinality $K-1$: $\mathcal{C}_{K-1}\triangleq \{\mathcal{K}\setminus \{k\}~|~k=1,\cdots,K\}$, and its cardinality is given by $|\mathcal{C}_{K-1}|=\big(\!\!\begin{tiny}\begin{array}{c}K\\K-1\end{array}\end{tiny}\!\!\big)=K$.

\item{Step 9:} Claim $K$ non-negative variables $\{x_{K-1,q}~|~q=1,2,\cdots,K\}$.

\item{Step 10:} Sum up all the variables $\{x_{K-1,q}~|~q=1,2,\cdots,K\}$ if the $q^{th}$ element in $\mathcal{C}_{K-1}$ contains $p$ for $p=1,\cdots,K$, and we upper bound them by non-negative $S_{(\mathcal{C}_{K-1})_p}$, respectively. That is,
\begin{eqnarray*}
\sum_{q=1}^Q x_{K-1,q}{\bf 1}(q\in (\mathcal{C}_{K-1})_p)\leq S_{(\mathcal{C}_{K-1})_p},~~~~p=1,2,\cdots,K.
\end{eqnarray*}
In addition, we also have $x_{K-1,q}\geq 0$, for $q=1,2,\cdots,K$, and $S_{(\mathcal{C}_{K-1})_p}\geq 0$. Add all these $3K$ inequalities to $\mathcal{E}$.

\item{Step 11:} Develop the cost function $\mathcal{L}^{K-1}$ as follows. If there is no coding, then we need a total of $\sum_{p=1}^K S_{(\mathcal{C}_{K-1})_p}$ transmissions. After we use coding,  we create a total of $\sum_{q=1}^K x_{K-1,q}$ equations, each involving $K-1$ intermediate symbols, and thus saving a total of $(K-2)(\sum_{q=1}^K x_{K-1,q})$ transmissions. Thus, the cost function is:
\begin{eqnarray*}
\mathcal{L}^{K-1}=\sum_{p=1}^K S_{(\mathcal{C}_{K-1})_p}-(K-2)\bigg(\sum_{q=1}^K x_{K-1,q}\bigg).
\end{eqnarray*}
Then we update the objective function as $\mathcal{L}=\mathcal{L}+\mathcal{L}^{K-1}$.

\item{Step 12:} Consider the set of variables $S_{(\mathcal{C}_j)_p}$ for every $p,~j$ where $p=1,2,\cdots,\big(\!\!\begin{tiny}\begin{array}{c}K\\j\end{array}\end{tiny}\!\!\big)$ and every $j\in\mathcal{K}$. We have the following constraints over the entire files:
\begin{eqnarray*}
\sum_{j=1}^K\sum_{p=1}^{\big(\!\!\begin{tiny}\begin{array}{c}K\\j\end{array}\end{tiny}\!\!\big)}S_{(\mathcal{C}_j)_p}=N,~~~~
\sum_{j=1}^K\sum_{p=1}^{\big(\!\!\begin{tiny}\begin{array}{c}K\\j\end{array}\end{tiny}\!\!\big)}jS_{(\mathcal{C}_j)_p}=\sum_{k=1}^K{M_k},
\end{eqnarray*}
and the following constraints over the files allocated at each server:
\begin{eqnarray*}
\sum_{j=1}^K\sum_{p=1}^{\big(\!\!\begin{tiny}\begin{array}{c}K\\j\end{array}\end{tiny}\!\!\big)}S_{(\mathcal{C}_j)_p}{\bf 1}(\textrm{If}~k~\textrm{is contained in the foot indices of} ~S_{(\mathcal{C}_j)_p})=M_k,~~~~\textrm{for}~\forall k\in\mathcal{K}.
\end{eqnarray*}
Add all these $K+2$ equations to $\mathcal{E}$.

\item{Step 13:} Finally, after removing redundant variables and constraints, we translate seeking the communication load problem into resolving the linear optimization problem:
\begin{eqnarray*}
\textrm{min} ~~ \mathcal{L},~~~~~~~~\textrm{subject to} ~~ \mathcal{E}.
\end{eqnarray*}

\item{Step 14:} We denote the optimal solution by $\mathcal{L}^o$, and the values of corresponding variables by $\mathcal{S}^o\triangleq \{S^o_{(\mathcal{C}_j)_p}~|~p=1,2,\cdots,\big(\!\!\begin{tiny}\begin{array}{c}K\\j\end{array}\end{tiny}\!\!\big),~j=1,2,\cdots,K\}$. According to $\mathcal{S}^o$, the corresponding file allocation $\mathcal{M}_k^o$ to each server $k$ for $k=1,2\cdots,K$ can be readily obtained.

\end{itemize}


\begin{remark}
It is worth noting that when $K$ increases, the number of variables and constraints grows much faster than $K$. When $K$ is large, even the linear optimization problem would be overwhelming, which prevents our algorithm from being applied for large $K$ due to the computational complexity. Therefore, an improved algorithm with lower complexity might be of interest in the future work.
\end{remark}

\section{Conclusion}

We investigate the MapReduce-based coded distributed computing (CDC) for heterogeneous systems by carefully designing file allocation and the optimal coding scheme to achieve the minimum communication load. While we completely resolve the minimum communication load for the system with $K=3$ by providing the achievability and the information theoretic converse, we provide an algorithm for the achievability for $K>4$. Future potential works would be to seek the minimum communication load for heterogeneous systems in the information theoretic sense under the MapReduced CDC framework.

\section{Acknowledgment}

This work is in part supported by ONR award N000141612189, NSA Award No. H98230-16-C-0255, a research gift from Intel. This material is also based upon work supported by Defense Advanced Research Projects Agency (DARPA) under Contract No. HR001117C0053. The views, opinions, and/or findings expressed are those of the author(s) and should not be interpreted as representing the official views or policies of the Department of Defense or the U.S. Government.

\bibliographystyle{IEEEtran}
{\footnotesize
\bibliography{heterogeneous_full}} 
\end{document}